%% file: Main.tex
\documentclass[twocolumn]{aastex631}
\usepackage{float}
\usepackage{booktabs}
\usepackage{graphicx}
\usepackage{amsmath}
\usepackage{xcolor}

\shorttitle{Broad H$\alpha$ Emitters in GOODS-N}
\shortauthors{Zhang et al.}

\graphicspath{{./}{figures/}}

\submitjournal{ApJ}

\begin{document}

\title{Abundant Population of Broad H$\alpha$ Emitters in the GOODS-N Field \\ Revealed by CONGRESS, FRESCO, and JADES}

\input{0_Author}

\begin{abstract}
We present a spectroscopic search for broad H$\alpha$ emitters at $z$\,$\approx$\,3.7--6.5 in the GOODS-N field, utilizing JWST/NIRCam slitless spectroscopy from FRESCO and CONGRESS, complemented by JADES imaging. We identify 19 broad H$\alpha$ emitters with FWHM $>$\,1000 km/s at $z$\,$\approx$\,4--5.5, including 9 new sources. The broad H$\alpha$ luminosity function (LF) derived from our sample is consistent with those of other JWST-selected broad-line AGN reported in the literature. The black hole masses and AGN bolometric luminosities, inferred from the broad H$\alpha$ components, indicate that most sources are accreting at $\sim$10\% of the Eddington limit. We derive their host stellar masses via SED fitting and find higher $M_{\rm BH}/M_{*}$ ratios relative to the local $M_{\rm BH}$–$M_\ast$ relations, consistent with previous studies. We find that 42\% of the sample do not satisfy the widely-used color selection criteria for Little Red Dots (LRDs), with the majority of these sources lacking the characteristic steep rest-optical red slope,
indicating that the LRD selection is highly incomplete when selecting AGN galaxies.
A comparison of the average SEDs between our sample and LRDs selected in the same field reveals that the steep red slopes observed in some LRDs are likely  due to line-boosting effects as previously suggested. Furthermore, we find that 68\% of color-selected LRDs with H$\alpha$ detections in the NIRCam/Grism spectra do not exhibit broad-line features. While the limited sensitivity of the grism spectra may hinder the detection of broad-line components in faint sources, our findings still highlight the enigmatic nature of the LRD population.

\end{abstract}

\keywords{AGN --- galaxies --- high-redshift --- supermassive black holes}

\input{1_Intro}
\input{2_Data}
\input{3_Selection}
\input{4_Results}
\input{5_Discussion}
\input{6_Conclusion}

\section*{Acknowledgments}

EE, YZ acknowledge support from the NIRCam Science Team contract to the University of Arizona, NAS5-02015. AJB acknowledges funding from the “FirstGalaxies” Advanced Grant from the European Research Council (ERC) under the European Union’s Horizon 2020 research and innovation program (Grant agreement No. 789056). RM acknowledges support by the Science and Technology Facilities Council (STFC), by the ERC through Advanced Grant 695671 “QUENCH”, and by the UKRI Frontier Research grant RISEandFALL. RM also acknowledges funding from a research professorship from the Royal Society. BER acknowledges support from the NIRCam Science Team contract to the University of Arizona, NAS5-02015, and JWST Program 3215. ST acknowledges support by the Royal Society Research Grant G125142. GV acknowledges support by European Union’s HE ERC Starting Grant No. 101040227 - WINGS. The research of CCW is supported by NOIRLab, which is managed by the Association of Universities for Research in Astronomy (AURA) under a cooperative agreement with the National Science Foundation.

This work is based on observations made with the NASA/ESA Hubble Space Telescope and NASA/ESA/CSA James Webb Space Telescope. The data were obtained from the Mikulski Archive for Space Telescopes at the Space Telescope Science Institute, which is operated by the Association of Universities for Research in Astronomy, Inc., under NASA contract NAS 5-03127 for JWST. These observations are associated with program \#1181 (JADES), 1895 (FRESCO), and 3577 (CONGRESS). 
The specific JWST observations analyzed can be accessed via \dataset[doi:10.17909/6rfk-6s81
]{http://dx.doi.org/10.17909/6rfk-6s81}.
Support for program \#3577 was provided by NASA through a grant from the Space Telescope Science Institute, which is operated by the Association of Universities for Research in Astronomy, Inc., under NASA contract NAS 5-03127.
The authors acknowledge the FRESCO team for developing their observing program with a zero-exclusive-access period.

\bibliography{Main}{}
\bibliographystyle{aasjournal}

\appendix
\section{Non-Confirmation of Two Broad H$\alpha$ Emitters Reported by Covelo-Paz et al.\ (2024)}
\label{appendix:badBLAGN}
\begin{figure}[htp]
    \centering
    \includegraphics[width=0.95\textwidth]{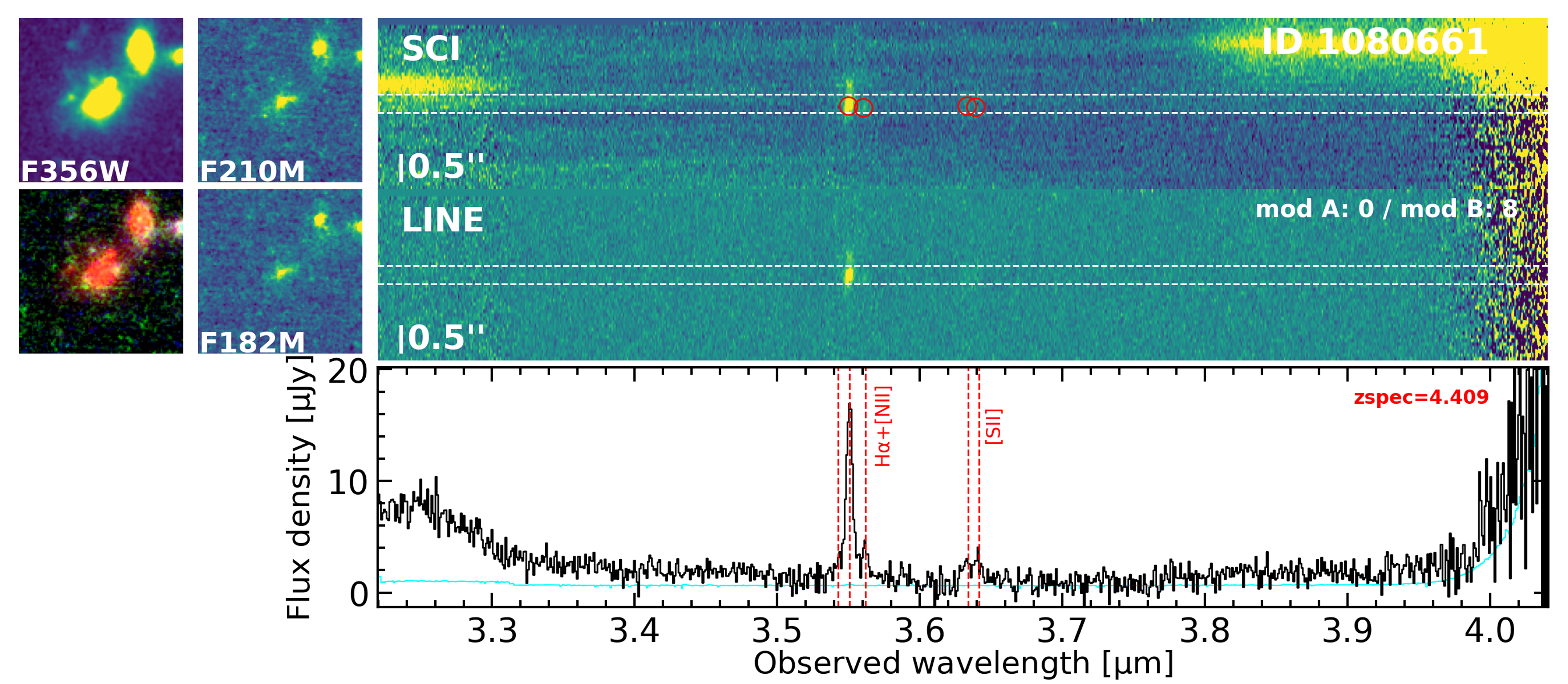}
    \caption{2D and 1D grism spectra of 1080661 obtained in the F356W filter from CONGRESS. Top row: 2D spectrum with continuum, red empty circles indicate detected lines. Second row: Line-only 2D spectrum. Third row: Optimally extracted 1D spectrum with continuum. Emission lines detected are marked by red dashed lines. The top-left corner displays the direct images of the source in the F182M, F210M, and F356W filters, along with a composite false-color stamp constructed from these three bands.}
    \label{figure: 1080661}
\end{figure}

\begin{figure}[htp]
    \centering
    \includegraphics[width=0.7\textwidth]{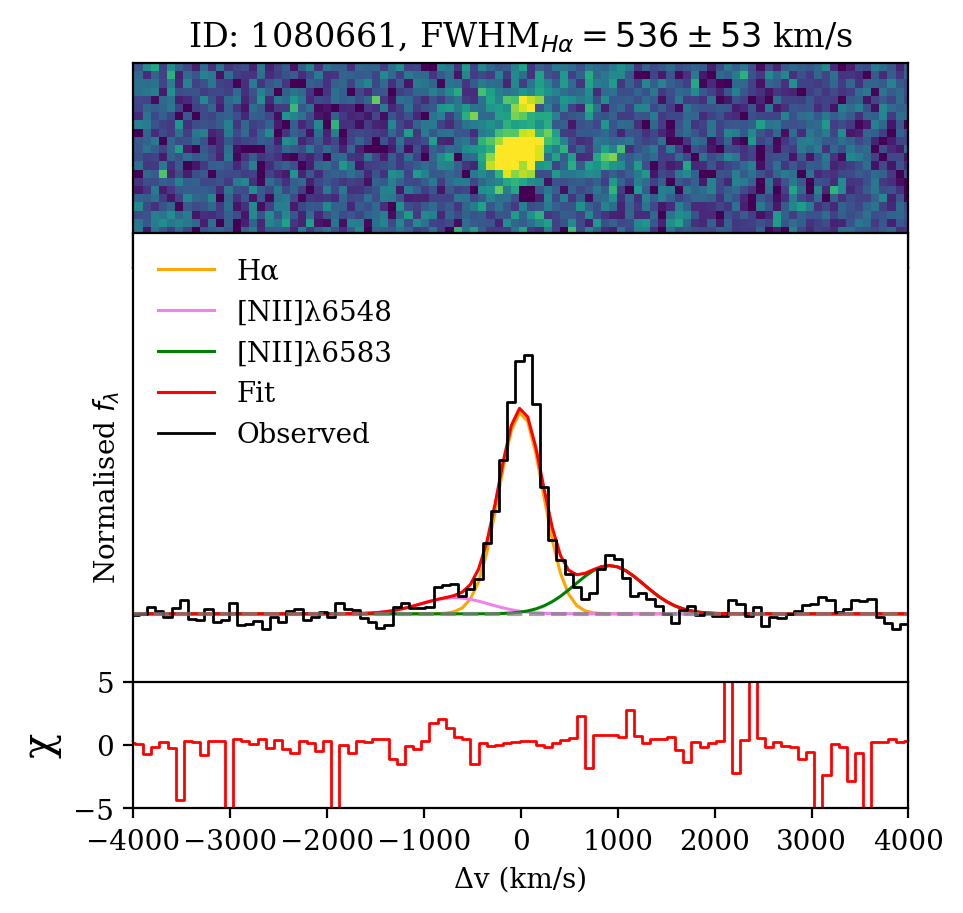}
    \caption{1D line fitting results for the H$\alpha$ emission profile of 1080661. Top panel: 2D grism spectrum with continuum subtracted. Middle panel: Optimally extracted 1D grism spectrum (black), while the red line represents the best-fit model. The yellow, pink, and green lines correspond to the H$\alpha$, [\ion{N}{2}]$\lambda6548$, and [\ion{N}{2}]$\lambda6583$ components, respectively. Bottom panel: residuals from the best-fit model.}
    \label{figure: 1080661_fit}
\end{figure}

In our sample, we exclude two broad H$\alpha$ emitters reported by \citet{Covelo_2024} due to concerns regarding the validity of their classification. For the first source (JADES ID: 1080661; RA = 189.27616 deg, Dec = 62.21416 deg), the flux ratio $\frac{F({\rm F444W}){r = 0.3''}}{F({\rm F444W}){r = 0.1''}} \approx 2.1$ does not meet our compactness criterion. Given that the selection criteria in \citet{Covelo_2024} do not incorporate compactness, this mismatch is natural. Figure~\ref{figure: 1080661} presents the 1D and 2D grism spectra obtained in F356W from CONGRESS. The F356W direct image reveals an extended morphology for this source. We also performed a 1D line fit of its H$\alpha$ line profile. As illustrated in Figure~\ref{figure: 1080661_fit}, even after accounting for potential morphological broadening, the best-fit FWHM is only $\sim$500 km/s, which does not satisfy our requirement for the line width of a broad H$\alpha$ emitter. Additionally, we note that our best-fit FWHM differs from the value reported in \citet{Covelo_2024} (i.e., 1674 km/s). However, the cause of this discrepancy is unclear.

\begin{figure}[htp]
    \centering
    \includegraphics[width=0.95\textwidth]{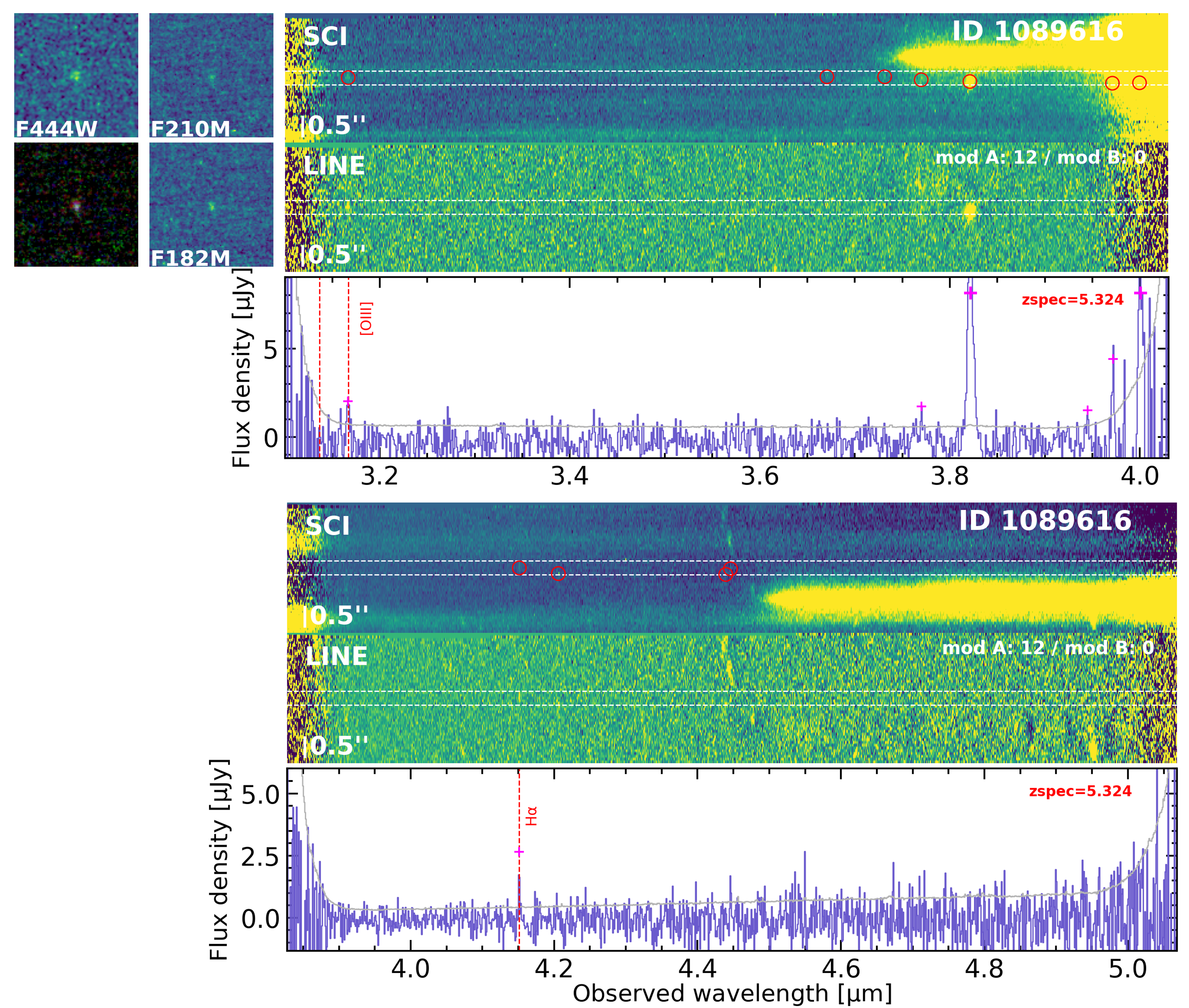}
    \caption{2D and 1D grism spectra of 1089616 obtained in the F356W (top) and F444W (bottom) filters from CONGRESS and FRESCO, respectively. Top row: 2D spectrum with continuum, red empty circles indicate detected lines. Second row: Median filtered line-only 2D spectrum. Third row: Optimally extracted 1D spectrum from the line-only 2D spectrum, with detected emission lines marked by red dashed lines. The top-left corner displays the direct images of the source in the F182M, F210M, and F444W filters, along with a composite false-color stamp constructed from these three bands.}
    \label{figure: 1089616}
\end{figure}

\begin{figure}[htp]
    \centering
    \includegraphics[width=0.95\textwidth]{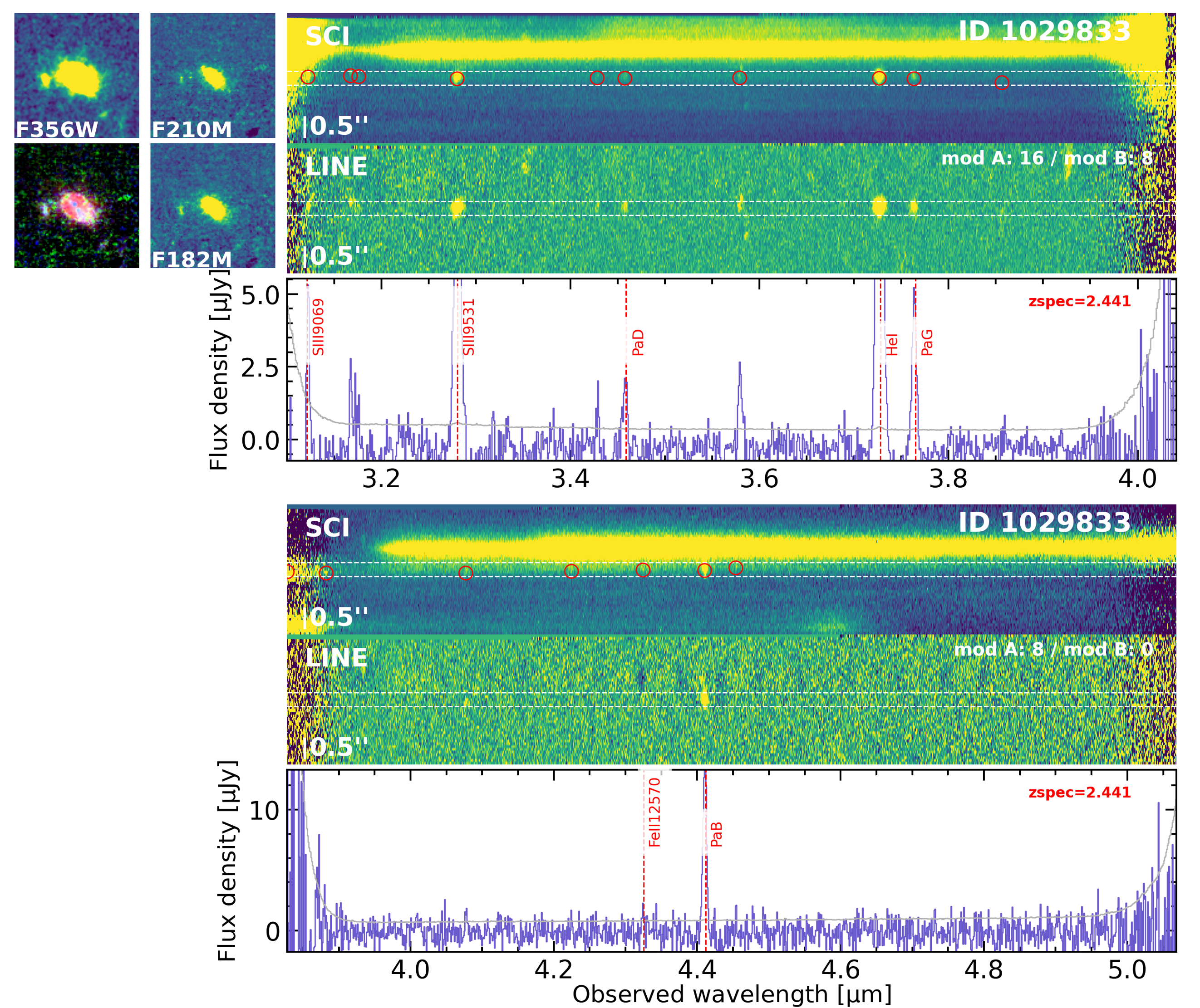}
    \caption{2D and 1D NIRCam/grism spectra of 1029833 obtained in the F356W (top) and F444W (bottom) filters from CONGRESS and FRESCO, respectively. Top row: 2D spectrum with continuum, red empty circles indicate detected lines. Second row: Median filtered line-only 2D spectrum. Third row: Optimally extracted 1D spectrum from the line-only 2D spectrum, with detected emission lines marked by red dashed lines. The top-left corner displays the direct images of the source in the F182M, F210M, and F444W filters, along with a composite false-color stamp constructed from these three bands.}
    \label{figure: 1029833}
\end{figure}

For the second source (JADES ID: 1089616; RA = 189.18713 deg, Dec = 62.27289 deg), the top panel of Figure~\ref{figure: 1089616} shows its 1D and 2D grism spectra in F356W from CONGRESS. While a strong emission feature is detected at 3.8 $\mu$m, we observe a noticeable spatial offset between the position of the emission line and that of the source in the direct image, raising concerns about its association with the target. In the bottom panel of Figure~\ref{figure: 1089616}, which presents the F444W grism spectra from FRESCO for the same source, an emission line at $\sim$4.15 $\mu$m is detected with no spatial offset and a S/N$\sim$4. Combined with the detection of another emission line at $\sim$3.17 $\mu$m, we identify these features as H$\alpha$ and [\ion{O}{3}]$\lambda5007$ of this source, respectively, confirming the redshift of the source is 5.324. The relatively weak line strengths are also consistent with the faint nature of the source in the direct images. Upon further inspection, the strong 3.8 $\mu$m line is confirmed to be [\ion{S}{3}]$\lambda9531$ from a nearby bright source (JADES ID: 1029833; RA = 189.20103 deg, Dec = 62.26567 deg). Figure \ref{figure: 1029833} presents the 1D and 2D grism spectra of this source in the F356W and F444W filters. The [\ion{S}{3}]$\lambda9531$ emission line of this source matches the 3.8~$\mu$m emission line observed in the 2D spectrum of source 1089616 very well, suggesting the 3.8~$\mu$m emission line is not H$\alpha$ of 1089616, but [\ion{S}{3}]$\lambda9531$ of 1029833. 

\section{Discrepant Broad-Line Widths for a BLAGN Reported by Greene et al.\ (2024)}

\label{appendix:prism}

\begin{figure}[!htbp]
    \centering
    \includegraphics[width=0.7\textwidth]{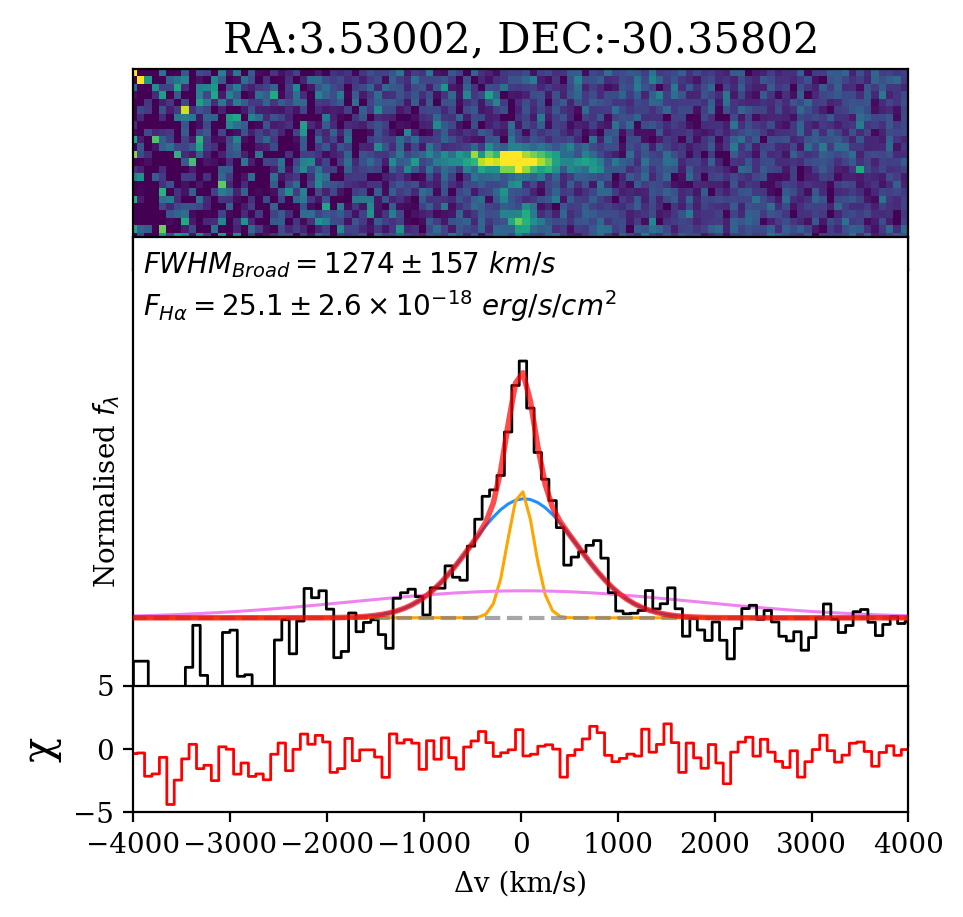}
    \caption{Multi-component Gaussian fit to the H$\alpha$ line profile in the NIRCam/grism spectrum (R$\sim$1600) for a BLAGN previously identified with the NIRSpec/prism in \citet{Greene_2024}. Top panel: 2D grism spectrum without continuum subtraction. Middle panel: Optimally extracted 1D spectrum (black), best-fit model (red), and individual components: narrow H$\alpha$ (yellow) and broad H$\alpha$ (blue). The pink line represents the broad component observed with the prism, using the FWHM and H$\alpha$ flux (treated as an upper limit of the flux for the broad component) reported by \citet{Greene_2024}. The intrinsic FWHM of the broad component and the observed H$\alpha$ flux are indicated in the top-left corner. Bottom panel: Residuals from the multi-component fit.}
    \label{figure: SAPHHIRES}
\end{figure}

Recently, the JWST Cycle 3 General Observer program Slitless Areal Pure-Parallel High-Redshift Emission Survey (SAPPHIRES-PID: 6434; PI: E. Egami; \citealt{Sun_2025}) revisited the Abell 2744 field and obtained NIRCam grism spectrum (R$\sim$1600) for one of the BLAGN (Prism ID: 38108; SAPPHIRES’s ID: 39082) identified in \citet{Greene_2024}. Figure \ref{figure: SAPHHIRES} shows the observed H$\alpha$ line in 1D and 2D grism spectra obtained in F444W filter, along with the line fitting results. The 1D grism spectrum reveals a broad H$\alpha$ component with a line width of $\sim$1300 km/s, which contrasts significantly with the prism-based value of $\sim$4000 km/s. For comparison, we also include the H$\alpha$ broad component observed in prism by using the FWHM and H$\alpha$ flux (treated as an upper limit of the flux for the broad component) reported by \citet{Greene_2024}. Even a modest difference in the FWHM measurements can result in a substantial variation in the inferred $M_{\rm BH}$. In Figure \ref{figure: LM}, we plot the AGN bolometric luminosity and black hole mass derived for this source from the SAPPHIRES NIRCam grism data (the ``X" marker).  With the smaller $M_{\rm BH}$ inferred from the grism spectrum, the source becomes more consistent with the distribution of our sample, providing a potential explanation for the discrepancy in black hole mass estimates. 

However, the underlying cause of the discrepancy in FWHM measurements remains unclear. Taking into account the large uncertainty associated with the prism-based FWHM, our measurement and that from \citet{Greene_2024} are broadly consistent with each other at the 2-$\sigma$ level. Since our line fitting was performed on spectra without continuum subtraction, we can rule out the possibility that the broad component was lost due to over-subtraction of continuum.  There are two possible explanations for this discrepancy. First, the difference in FWHM values may arise from differences in the respective LSFs. Second, the grism observation may lack the sensitivity required to detect such a faint broad component. Furthermore, we note that \citet{Greene_2024} identified broad-line sources using slightly different selection criteria, in particular requiring the FWHM of the broad H$\alpha$ component to $>$ 2000 km~s$^{-1}$. This difference in criteria may also contribute to the discrepancy in FWHM measurements.



\section{NIRSpec Medium Grating vs. NIRCam Grism}
\label{appendix:grating}
\begin{figure}[htp]
    \centering
    \includegraphics[width=0.95\textwidth]{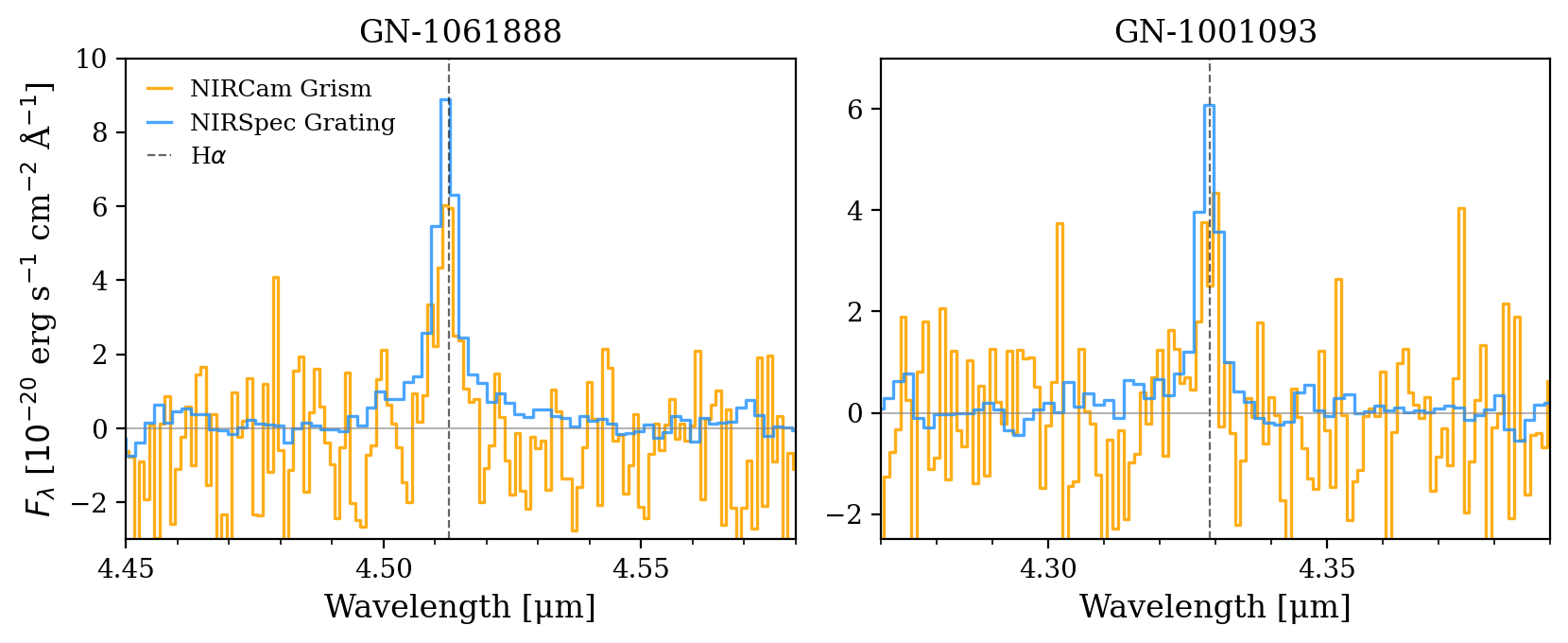}
    \caption{Comparison of the H$\alpha$ line profiles for two LRDs identified in \citet{Pier_2024}, as observed from NIRCam grism (orange) and NIRSpec grating (blue) spectra. The dashed gray line marks the centroid wavelength of H$\alpha$.}
    \label{figure:appendix}
\end{figure}

In this work, we identify two photometrically selected LRDs from \citet{Pier_2024} that exhibit broad H$\alpha$ line profiles in the NIRSpec medium resolution grating spectra presented by \citet{Maiolino_2024a}, but not in the NIRCam grism spectra, likely due to the limited sensitivity of the grism observations. Figure \ref{figure:appendix} illustrates the differences in the observed H$\alpha$ line profiles for these two sources between the grism and grating data. The S/N of the grating spectra are slightly lower than those presented in \citet{Maiolino_2024a}, possibly due to differences in the spectra extraction methods. In both cases, the H$\alpha$ lines appear significantly narrower in the grism spectra compared to the grating spectra. Additionally, the grism continuum exhibits a much higher noise level, supporting the possibility that the broad components are undetected due to the lower S/N of the grism observations.

\end{document}

%% file: 0_Author.tex
\author[0000-0002-1574-2045]{Junyu Zhang}
\affiliation{Steward Observatory, University of Arizona, 933 N Cherry Ave, Tucson, AZ 85721, USA}

\author[0000-0003-1344-9475]{Eiichi Egami}
\affiliation{Steward Observatory, University of Arizona, 933 N Cherry Ave, Tucson, AZ 85721, USA}

\author[0000-0002-4622-6617]{Fengwu Sun}
\affiliation{Center for Astrophysics $|$ Harvard \& Smithsonian, 60 Garden St., Cambridge, MA 02138, USA}

\author[0000-0001-6052-4234]{Xiaojing Lin}
\affiliation{Steward Observatory, University of Arizona, 933 N Cherry Ave, Tucson, AZ 85721, USA}
\affiliation{Department of Astronomy, Tsinghua University, Beijing 100084, China}

\author[0000-0002-6221-1829]{Jianwei Lyu}
\affiliation{Steward Observatory, University of Arizona, 933 N Cherry Ave, Tucson, AZ 85721, USA}

\author[0000-0003-3307-7525]{Yongda Zhu}
\affiliation{Steward Observatory, University of Arizona, 933 N Cherry Ave, Tucson, AZ 85721, USA}

\author[0000-0002-5104-8245]{Pierluigi Rinaldi}
\affiliation{Space Telescope Science Institute, 3700 San Martin Drive, Baltimore, Maryland 21218, USA}

\author[0000-0001-6561-9443]{Yang Sun}
\affiliation{Steward Observatory, University of Arizona, 933 N Cherry Ave, Tucson, AZ 85721, USA}

\author[0000-0002-8651-9879]{Andrew J.\ Bunker }
\affiliation{Department of Physics, University of Oxford, Denys Wilkinson Building, Keble Road, Oxford OX1 3RH, UK}

\author[0000-0003-0883-2226]{Rachana Bhatawdekar}
\affiliation{European Space Agency (ESA), European Space Astronomy Centre (ESAC), Camino Bajo del Castillo s/n, 28692 Villanueva de la Cañada, Madrid, Spain}

\author[0000-0003-4337-6211]{Jakob M.\ Helton}
\affiliation{Department of Astronomy \& Astrophysics, The Pennsylvania State University, University Park, PA 16802, USA}

\author[0000-0002-4985-3819]{Roberto Maiolino}
\affiliation{Kavli Institute for Cosmology, University of Cambridge, Madingley Road, Cambridge, CB3 0HA, UK}
\affiliation{Cavendish Laboratory - Astrophysics Group, University of Cambridge, 19 JJ Thomson Avenue, Cambridge, CB3 0HE, UK} 
\affiliation{Department of Physics and Astronomy, University College London, Gower Street, London WC1E 6BT, UK}

\author[0009-0003-5402-4809]{Zheng Ma}
\affiliation{Steward Observatory, University of Arizona, 933 N Cherry Ave, Tucson, AZ 85721, USA}

\author[0000-0002-4271-0364]{Brant Robertson}
\affiliation{Department of Astronomy and Astrophysics University of California, Santa Cruz, 1156 High Street, Santa Cruz CA 96054, USA} 

\author[0000-0002-8224-4505]{Sandro Tacchella}
\affiliation{Kavli Institute for Cosmology, University of Cambridge, Madingley Road, Cambridge, CB3 0HA, UK}
\affiliation{Cavendish Laboratory, University of Cambridge, 19 JJ Thomson Avenue, Cambridge, CB3 0HE, UK}

\author[0000-0001-8349-3055]{Giacomo Venturi}
\affiliation{Scuola Normale Superiore, Piazza dei Cavalieri 7, I-56126 Pisa, Italy}

\author[0000-0003-2919-7495]{Christina C.\ Williams}
\affiliation{NSF National Optical-Infrared Astronomy Research Laboratory, 950 North Cherry Avenue, Tucson, AZ 85719, USA}

\author[0000-0002-4201-7367]{Chris Willott}
\affiliation{NRC Herzberg, 5071 West Saanich Rd, Victoria, BC V9E 2E7, Canada}

%% file: 1_Intro.tex
\section{Introduction} \label{sec:intro}
Supermassive black holes (SMBHs) are thought to reside at the centers of most massive galaxies \citep{Kormendy_2013, Greene_2020}. The efficient accretion processes associated with these SMBHs contribute to the formation of active galactic nuclei (AGN), causing their host galaxies to exhibit broadened emission line profiles compared to non-active galaxies. Understanding the characteristics of AGN and accurately estimating their abundance in the early Universe ($z > 3$) are essential for addressing key questions in extragalactic astronomy, including the reionization of the Universe and the co-evolution of AGN and their host galaxies. 

Prior to the launch of the JWST \citep{Gardner_2023}, studying the abundance and properties of faint AGN at high redshifts was highly challenging due to observational limitations. For AGN with ultraviolet (UV) magnitudes fainter than $M_{\text{UV}} = -22$, determining their space density and physical properties was particularly difficult, with significant discrepancies reported \citep[e.g.,][]{Parsa_2018, Giall_2019, Morishita_2020, Shen_2020, Fink_2022}. The advent of JWST, with its unparalleled infrared sensitivity, has opened new avenues for investigating these faint, distant objects. Recent JWST observations have demonstrated its remarkable capability to identify UV-faint AGN in the early Universe through various methodologies, including the detection of broad Balmer lines \citep{Kocevski_2023, Harikane_2023, Oesch_fresco, Larson_2023, ubler_2023, Barro_2024, Matthee_2024, Maiolino_2024a, Greene_2024, Lin_LRD}.

Among the UV-faint AGN detected by JWST, an intriguing subset of the AGN population consists of sources with notably compact sizes and red colors. \citet{Matthee_2024} assembled one of the earliest samples of these sources, identified through broad H$\alpha$ emission lines, and named them as ``little red dots" (LRDs). These LRDs appear as point sources and exhibit distinctive spectral energy distributions (SEDs), characterized by blue rest-frame UV slopes and extremely steep red rest-frame optical slopes \cite[e.g.,][]{Greene_2024, Labbe_2025, Ji_2025, Naidu_2025, Graaff_2025}. Such a ``V-shaped" SED is found to be rare for AGN at $z <$ 2 (e.g., \citealt{Nobori_2019}). To investigate the origin of the UV and optical emission in LRDs, several recent studies have assembled samples of these objects by identifying compact sources and targeting those with ``V-shaped" SEDs, using NIRCam color cuts as a selection criterion \citep{Akins_2023, Barro_2024, Greene_2024, Pablo_2024, Kokorev_2024, Williams_2024,Pier_2024, Labbe_2025}.

Although numerous photometrically selected LRD candidates are assembled, their true nature remains uncertain. It is still unclear whether these LRDs are primarily AGN, in which the red optical continua arise from dust emission in front of UV-bright AGN \citep[e.g.,][]{Akins_2023, Maiolino_2024a} or non-stellar Balmer breaks \citep{Lin_2025b}, or whether they are dusty star-forming galaxies, where strong emission lines such as H$\alpha$ or [\ion{O}{3}] with high equivalent widths can boost the broad-band fluxes and make the rest-frame optical colors appear red \citep[e.g.,][]{Barro_2024, Pablo_2024, Hviding_2025}. An additional possibility is that some LRDs may be ``black hole stars” embedded in relatively bright host galaxies, in which extremely dense gas forms a dust-free atmosphere around a SMBH, producing strong Balmer break and absorption features \citep{Naidu_2025}. \citet{Greene_2024} obtained follow-up NIRSpec spectroscopy on 15 LRDs photometrically selected in \citet{Labbe_2025} and found that 60\% exhibited definitive evidence of broad H$\alpha$ lines, confirming their AGN nature for a significant subset. In contrast, \citet{Pablo_2024} reported that only 17\% of their photometrically selected LRDs with available spectroscopy displayed broad line features. However, this fraction may be underestimated, as many existing broad lines could not be confirmed due to insufficient S/N of the grism spectra \citep{Williams_2024}. In \citet{Pier_2024}, analysis of the emission line properties of the LRD sample using NIRSpec spectra and classical line ratio diagnostics was unable to conclusively determine their nature, suggesting a likely mixed origin.

Furthermore, SED-fitting results for these LRDs indicate that the origin of their red colors remains unclear. Although SED fits to LRDs with JWST and ALMA observations favor models with dusty AGN over obscured star formation to explain their red optical colors, this does not completely rule out a stellar-dominated origin, as dust-obscured older stellar populations remain consistent with the ALMA limits \citep{Williams_2024,Labbe_2025}.

In this work, we utilized grism spectroscopic data from the First Reionization Epoch Spectroscopically Complete Observations (FRESCO-PID: 1895; PI: P. Oesch; \citealt{Oesch_fresco}) and the Complete NIRCam Grism Redshift Survey (CONGRESS-PID: 3577; PI: E. Egami; Sun et al., in prep, \citealt{Lin_congress}), along with imaging data from the JWST Advanced Deep Extragalactic Survey (JADES-PID: 1181; PI: D. Eisenstein; \citealt{JADES, JADES_doi}), to conduct a comprehensive search for broad H$\alpha$ emitters at $z\approx 3.7-6.5$ in the GOODS-North field (\citealt{Giavalisco_2004}, hereafter GOODS-N). After identifying this sample of broad H$\alpha$ emitters, we aim to characterize their physical properties and compare them with LRDs recently identified in other studies. The environments of these AGNs and their impact on BH growth are discussed in a companion paper \citep{Lin_2025a}.

This paper is organized as follows: In Section~\ref{sec: data}, we provide a brief overview of the spectroscopic and imaging data utilized in this study. Section~\ref{sec: selection} details the selection process for broad H$\alpha$ emitters and the methodology for 1D line fitting. In Section~\ref{sec: results}, we present and discuss the results of the 1D line fitting for our broad H$\alpha$ emitters. Section~\ref{sec: Disc} explores the physical properties of the broad H$\alpha$ emitter sample and compares our sample with the LRD sample identified in the same field. Finally, Section~\ref{sec:highlight} summarizes the main findings of this work.

Throughout this paper, we assume a flat $\Lambda$CDM cosmology with $\Omega_{m}$ = 0.3 and $H_0 = 70$ km/s/Mpc. Magnitudes are listed in the AB system \citep{Oke_AB}.

%% file: 2_Data.tex
\section{Data} \label{sec: data}
\subsection{Spectroscopic Data} \label{subsec: specdata}
The sample of broad H$\alpha$ emitters analyzed in this paper has been selected based on the spectroscopic data obtained through the JWST/NIRCam Wide
Field Slitless Spectroscopy (WFSS) mode with the FRESCO and CONGRESS programs. The FRESCO survey covers 62~\sq{\arcmin} in each of the two GOODS fields, acquiring about 2-hour deep NIRCam/grism observations using the F444W filter. These observations provide grism spectra with a resolving power of $R\sim1600$, covering $3.8–5.0$\,\micron\ for most galaxies in the field-of-view (FOV). The CONGRESS covers the GOODS-N field only, with a footprint nearly identical to that of FRESCO. The observations with the F356W filter yield grism spectra with a resolving power of $R\approx1400-1610$  from 3.1 to 4.0\,\micron. Since the GOODS-S field has grism spectroscopic coverage only in F444W, and \citet{Matthee_2024} have already performed a systematic search for broad H$\alpha$ emitters using these data, this study focuses on the GOODS-N field only. By combining the spectroscopic data from FRESCO and CONGRESS, we are able to search for broad H$\alpha$ emitters within a redshift range of $z=3.7-6.5$ in the GOODS-N field.

All the NIRCam/WFSS data were reduced using the publicly available reduction pipeline described in \citet{Sun_grism}. Here, we briefly summarize the main process. The WFSS data were first reduced to the stage-1 level by using the standard JWST stage-1 calibration pipeline \texttt{v1.11.2}. We then applied a pixel-to-pixel flat-field correction, performed the 2D sky-background subtraction, and assigned world coordinate system (WCS) information for each exposure. The WCS of the grism exposures was calibrated with the Gaia DR3 catalog \citep{gaia_dr3} by cross-matching stars identified in the simultaneously acquired NIRCam short-wavelength images with the corresponding sources in the Gaia catalog. For each source, we utilized the grism spectral tracing and dispersion models, along with the flux calibration functions presented in \citet{Sun_grism}, to extract the 2D spectra from individual grism exposures. The extracted spectra were then coadded in a common wavelength and spatial grid. Finally, we extracted the 1D spectra using optimal extraction algorithms \citep{Horne1986} and applied median filters to subtract the background and the continuum.

\subsection{Multi-band Imaging Data}\label{subsec: image}
To obtain the morphology information and estimate the physical properties of our broad H$\alpha$ emitters, we utilize multi-band imaging data from JADES. For the GOODS-N field, 9 NIRCam images with the following filters are available: F090W, F115W, F150W, F200W, F277W, F335M, F356W, F410M, and F444W, covering a wide range of wavelengths. However, JADES covers the GOODS-N field with a different footprint compared to those of FRESCO and CONGRESS. The total overlapping area between JADES and FRESCO/CONGRESS footprints is $\sim35$~\sq{\arcmin} in GOODS-N. Therefore, for about half of our broad H$\alpha$ emitters, which are only covered by FRESCO and CONGRESS (see Figure \ref{figure: fp}), imaging data are available in only six filters: F090W, F115W, F182M, F210M, F356W, and F444W. Additionally, for each broad H$\alpha$ emitter, we also checked whether deep imaging data in F435W, F606W, F775W, F814W, and F850LP are available from HST/ACS and HST/WFC3 (e.g., \citealt{Koek_2011}). Overall, for our broad H$\alpha$ emitters with $z\approx3.9-5.5$, these multi-band imaging data provide excellent coverage of the rest-frame wavelengths from UV to optical (i.e., $< 1~\micron$), which allows us to constrain the physical properties of these broad H$\alpha$ emitters through SED-fitting.

%% file: 3_Selection.tex
\section{Broad H\texorpdfstring{$\alpha$}{Lg} Emitter Selection} \label{sec: selection}

Given the large FOV of FRESCO and CONGRESS, conducting a blind search for broad H$\alpha$ emitters using the grism spectra across the field is extremely time-consuming. Additionally, an extended morphology can produce a fake broad-line feature in the grism data. Therefore, to identify broad H$\alpha$ emitters more efficiently, we first construct a galaxy sample by applying compactness and photometric redshift (photo-$z$) criteria. Specifically, we require each source to have $\frac{F(F444W){r = 0.3''}}{F(F444W){r = 0.1''}} < 1.2$ after applying aperture correction, to ensure the selection of compact sources. To further reduce the sample size, we only include sources with photo-$z$ values reported in JADES DR2 \citep{Eugenio_2025} between 3 and 7, which is slightly broader than the redshift range where the H$\alpha$ emission line is expected to be observable in the grism spectra from FRESCO/CONGRESS, accounting for the uncertainty in photo-$z$. In this way, we obtained a sample of $\sim600$ galaxies, which is manageable for visual inspection. 

Next, we conducted a visual investigation of the 1D grism spectra for each galaxy to remove contaminants that do not show a clear H$\alpha$ emission line, either because no line is detected or because the observed line is inconsistent with H$\alpha$. For those potential broad H$\alpha$ emitters identified by visual inspection, we fit their emission-line profiles with a multi-component Gaussian profile to determine the line widths of their broad components (see Section \ref{sec: linefit} for more details). To select sources with broad-line feature, we require our broad H$\alpha$ emitters to have a broad component with a full width at half maximum (FWHM$_{\text{H}\alpha, \text{broad}}$) greater than 1000 km/s. We retain only objects in which the broad component is detected with S/N $>$ 3 in line flux to ensure the broad component is real. The FWHM of the [\ion{O}{3}] doublet is also measured when available. Below, we summarize our selection criteria:
\begin{equation} \label{eq1}
\begin{split}
 \frac{F({\rm F444W}){r = 0.3''}}{F({\rm F444W}){r = 0.1''}} & < 1.2 \\
 3 < z_{\rm photo} & < 7 \\
{\rm S/N_{H\alpha,~broad}} & > 3 \\
{\rm FWHM_{H\alpha,~broad}}&  > 1000~\text{km/s} 
\end{split}
\end{equation}

\subsection{1D Line Fitting} \label{sec: linefit}
To determine the line widths of the broad component in these broad H$\alpha$ emitters, we performed a simultaneous fit of both the narrow and broad components for H$\alpha$, and a narrow-component only fit for [\ion{N}{2}]$_{6549,6585}$. The line ratio of the [\ion{N}{2}] doublet was fixed to 1:3.049, in accordance with recent results from \citet{Doj_2023}. Assuming that both the narrow component of H$\alpha$ and the [\ion{N}{2}] doublet originate from the narrow line region, we constrained their line centroids and line widths to be identical. However, as the centroid of the broad H$\alpha$ component is sometimes observed to deviate slightly from that of the narrow component (e.g., \citealt{Matthee_2024}), we allow the broad and narrow components to have different line centroids by $\pm$0.01 \micron~(corresponding to $\approx600–1000$\,km/s, depending on the redshift).

In addition to fitting the line profile using a straightforward combination of narrow and broad Gaussian components, we also tested a model incorporating H$\alpha$ absorption, although previous studies suggest that detecting such absorption features requires high S/N (e.g., \citealt{Juodzbalis_2024, Ji_2025, DEugenio_2025, Rusakov_2025}). To robustly confirm the presence of H$\alpha$ absorption, we roughly follow the methodology outlined in \citet{Maiolino_2024a} and perform a Bayesian Information Criterion (BIC) test \citep{Liddle_2007}. For Gaussian noise, the BIC parameter is defined as:
\begin{equation}
    BIC = \chi^2 + k\ln{n},
\end{equation}
where $k$ is the number of free parameters in the model and $n$ is the number of data points used to fit with. We require the BIC value of the fit with H$\alpha$ absorption to be at least a factor of 6 larger than that of the fit without it:
\begin{equation}
    BIC_{\rm No~Absorption} - BIC_{\rm Absorption} > 6
\end{equation}

%% file: 4_Results.tex
\section{Results} \label{sec: results}

\subsection{19 Broad H\texorpdfstring{$\alpha$}{Lg} Emitters at \texorpdfstring{$z=$}{Lg}3.9-5.5}

\begin{figure*}[!htp]
    \centering
    \includegraphics[width=0.95\textwidth]{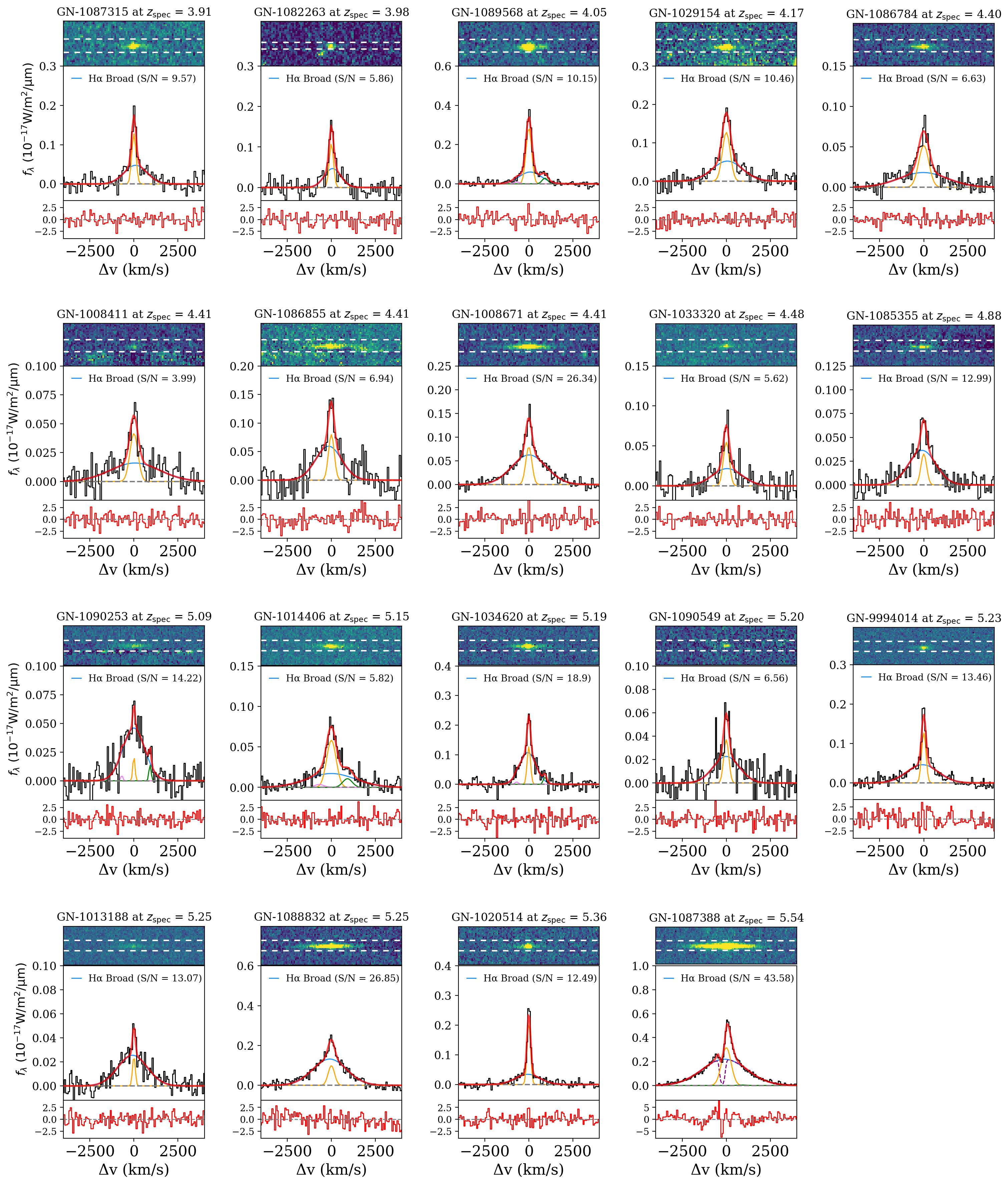}
    \caption{Multi-component Gaussian fits to the H$\alpha$ line profile for the 19 broad H$\alpha$ emitters. Top panels: Continuum-subtracted 2D spectrum. The white dashed lines represent the size of aperture used to extract 1D spectrum for each source. Middle panels: Optimally extracted 1D spectrum (black), best-fit model (red), and individual components (narrow H$\alpha$: yellow; broad H$\alpha$: blue; [\ion{N}{2}] doublet: pink and green). Bottom panels: Residuals from the multi-component model.}
    \label{figure: fit}
\end{figure*}

We have identified 19 broad H$\alpha$ emitters in the GOODS-N field, with 10 originating from CONGRESS and 9 from FRESCO. Among the 19 sources, 9 are new discoveries in this work. Figure \ref{figure: fit} shows the line fitting results for all 19 broad H$\alpha$ emitters. Tables \ref{table:Congress} and \ref{table:Fresco} summarize the basic information and line fitting results for the broad H$\alpha$ emitters identified in CONGRESS ($3.91 < z < 4.88$) and FRESCO ($5.09 < z < 5.54$), respectively. The reported FWHMs represent the intrinsic FWHMs of the emitters, de-convolved by the line spread function (LSF) as described in Sun et al. (in prep).

\begin{figure*}[!htpb]
    \centering
    \includegraphics[width=0.95\textwidth]{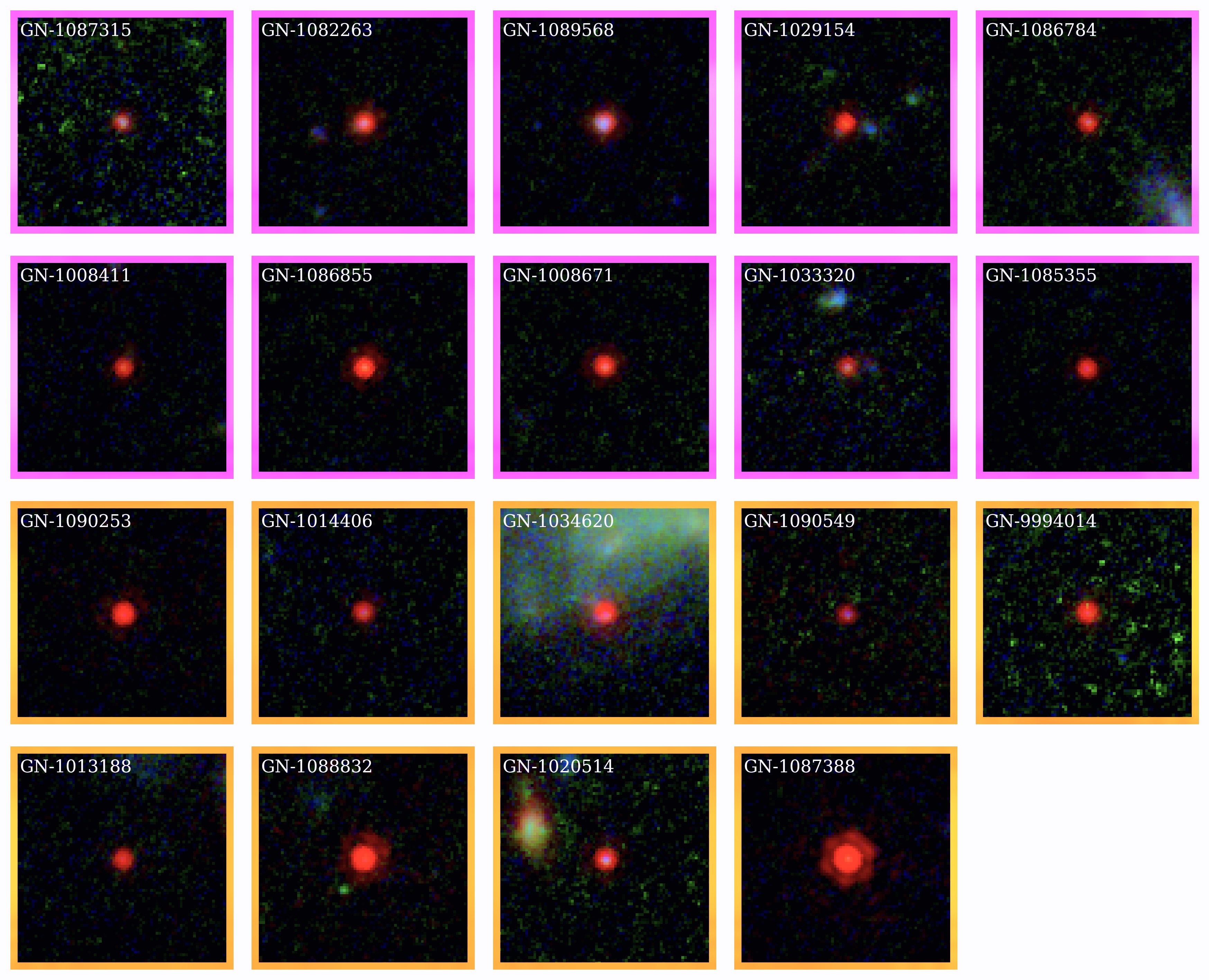}
    \caption{False-color stamps of the 19 broad H$\alpha$ emitters identified in this work. Emitters detected in FRESCO are marked in orange and the stamps are constructed using NIRCam F115W+F210M+F444W imaging data, while those detected in CONGRESS are marked in violet and the stamps are constructed using NIRCam F115W+F210M+F356W imaging data. Each stamp spans 2.5$\times$2.5 \sq\arcsec.}
    \label{figure: rgb}
\end{figure*}

\begin{figure}[!hptb]
    \centering
    \includegraphics[width=0.49\textwidth]{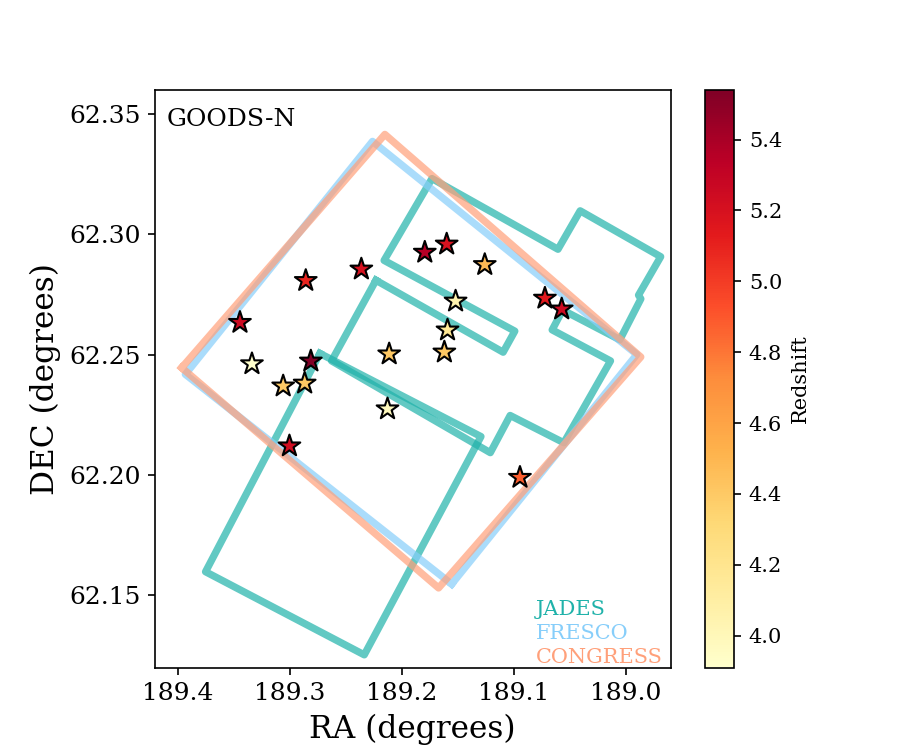}
    \caption{Sky positions of the 19 broad H$\alpha$ emitters (stars) discovered in the GOODS-N field, color-coded by redshift. The JADES, FRESCO, and CONGRESS footprints are overlaid with light green, blue, and red lines, respectively.}
    \label{figure: fp}
\end{figure}

Among all broad H$\alpha$ emitters, the broad H$\alpha$ components exhibit average FWHM of 2032 km/s (ranging from 1077--3281 km/s) and luminosity of 4.40$\times10^{42}$ erg/s (ranging from 0.9--32.9$\times10^{42}$ erg/s). [\ion{N}{2}] emission with a S/N$>$3 is detected in only two objects (i.e., GN-1034620 and GN-1089568). H$\alpha$ absorption is also found to be rare within our sample, with a robust detection observed in only one object (i.e., GN-1087388), consistent with \citet{Matthee_2024}, who also reported the presence of an absorption feature in this source. However, we cannot assess the presence of H$\alpha$ absorption with grism data. As noted earlier, recent studies with NIRSpec have shown that detecting such absorption features typically requires high S/N or higher resolution (e.g., \citealt{Juodzbalis_2024, Ji_2025, DEugenio_2025, Rusakov_2025}). Given the relatively low S/N of many sources in our sample, it remains possible that H$\alpha$ absorption is present but remains undetected in some cases.

Although the line centroids of the broad and narrow components are allowed to be different, the majority of objects exhibits offsets of $<$ 100 km/s, with the largest offset of $150$\,km/s observed in GN-1086855. We find that the broad components contribute, on average, about 72\% of the total H$\alpha$ luminosity, with only one object exhibiting a broad component contribution of less than 50\%. We note that our sample exhibits a higher average contribution from the broad component compared to that of the Type 1 AGN identified with NIRSpec in JADES (i.e., 46\% $\pm$ 25\%; \citealt{Juodzbalis_2025}). This difference is likely due to the lower S/N of our sample, which may hinder the detection of faint broad components. However, by taking the uncertainties into consideration, our average broad-component contribution remains consistent with that of \citet{Juodzbalis_2025} at the 1-$\sigma$ level.

Figure \ref{figure: rgb} presents the false-color stamp images of the 19 discovered broad H$\alpha$ emitters while Figure \ref{figure: fp} illustrates the sky positions of these emitters, overlaid with the footprints of CONGRESS, FRESCO, and JADES. Notably, our sample includes 7 broad H$\alpha$ emitters from FRESCO, as reported by \citet{Matthee_2024}, and 3 broad H$\alpha$ emitters from CONGRESS, as reported by \citet{Covelo_2024}. We note that two broad H$\alpha$ emitters in \citet{Covelo_2024} were excluded as they do not meet our selection criteria. A detailed explanation is provided in Appendix~\ref{appendix:badBLAGN}. 

We highlight 9 broad H$\alpha$ emitters newly identified in this work in Table \ref{table:Congress} and \ref{table:Fresco}. The exact reasons for their absence in previous studies remain unclear.  For the seven CONGRESS sources not repoted in \citet{Covelo_2024}, the paper does not provide enough information to make assessment.
Regarding the two FRESCO sources not reported in \citet{Matthee_2024}, we suspect that GN-1090549 may have been excluded due to its relatively low S/N, as \citet{Matthee_2024} adopted a stricter cut (S/N $>$ 5) than we did. Although GN-1090549 shows a broad component with S/N $>$ 5 in our analysis, differences in 1D grism spectra extraction methods between the two studies may explain this discrepancy. For GN-1034620, we note that this source overlaps with a giant local galaxy, and it is possible that segmentation differences led to it being excluded from their sample.

\subsection{Narrow [\texorpdfstring{\ion{O}{3}}{Lg}] Lines and AGN Origin of the Broad H\texorpdfstring{$\alpha$}{Lg} Lines}

\begin{figure*}[!htp]
    \centering
    \includegraphics[width=0.95\textwidth]{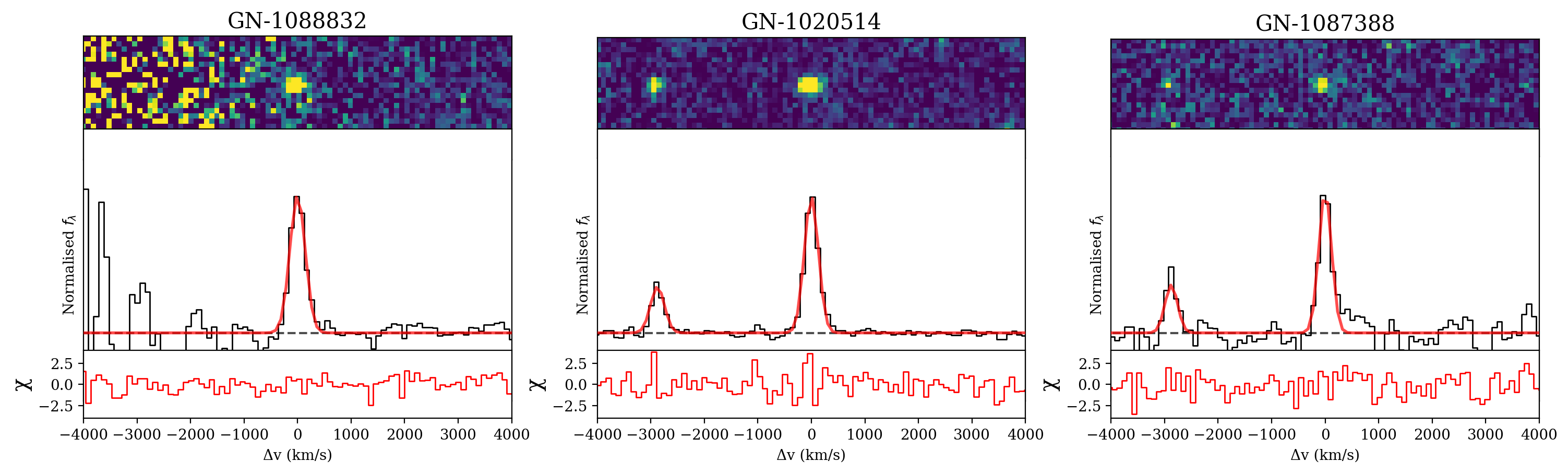}
    \caption{Single-component Gaussian fits (red) to the observed [\ion{O}{3}] line profiles (black) of 3 broad H$\alpha$ line emitters. Top panel: Continuum-subtracted 2D spectrum. Bottom panel: Residuals from the single-component Gaussian model. Line-fitting results are summarized in Table \ref{table:Fresco}.}
    \label{figure: o3fit}
\end{figure*}

Since the JWST/NIRCam grism spectra in our study only cover two wideband filters (i.e., F356W and F444W), the [\ion{O}{3}] doublet is within the wavelength coverage for only three emitters in our sample. Figure \ref{figure: o3fit} shows the 1D line fitting results for these three sources. Since the FWHM observed in [\ion{O}{3}] is significantly smaller than that observed in H$\alpha$, this effectively rules out the possibility that the broad components originate from galactic outflows.

For emitters without [\ion{O}{3}] coverage, although we cannot entirely exclude the possibility of galactic outflows, the majority exhibit broad component FWHM values significantly larger than 1000 km/s, along with broad-to-narrow H$\alpha$ flux ratios exceeding 50\%. In contrast, broad components associated with galactic outflows typically exhibit narrower FWHMs, ranging from a few hundred to $\sim$1000 km/s, depending on whether the outflows are driven by star formation or AGN activity, and show broad-to-narrow H$\alpha$ flux ratios of $\sim$50\% (see \citealt{Matthee_2024} and references therein). We also note that in published NIRSpec studies, nearly all broad H$\alpha$ emitters show narrow [\ion{O}{3}] emission, with their broad Balmer lines consistently attributed to the AGN broad line region (BLR) (e.g., \citealt{Harikane_2023, Greene_2024}). Therefore, it is reasonable to assume that the broad components in our sample are primarily caused by AGN activity.

\input{Tables/Tab1_Congress}

\input{Tables/Tab2_Fresco}

%% file: Tables/Tab1_Congress.tex
\begin{deluxetable*}{lcccccc}
\tablewidth{0pt}
\caption{Line-fitting results for the 10 broad H$\alpha$ emitters identified from CONGRESS.}
\label{table:Congress}
\tablehead{
\colhead{JADES ID\tablenotemark{a}} & \colhead{RA} & \colhead{DEC} & \colhead{$z_{\rm spec}$} & \colhead{$f_{\rm H\alpha}$} & \colhead{FWHM$_{\rm H\alpha,~Broad}$} & \colhead{F356W} \\
\colhead{} & \colhead{(deg)} & \colhead{(deg)} & \colhead{} & \colhead{($10^{-21}$ W/m$^2$)} & \colhead{(km/s)} & \colhead{(mag)} 
}
\startdata
GN-1087315\tablenotemark{b} & 189.333584 & 62.246178 & 3.91 & 12.5$\pm$1.0 & 1514$\pm$183 & 25.73 \\
GN-1082263\tablenotemark{b} & 189.212584 & 62.227436 & 3.98 &\phn 9.6$\pm$1.3&  1083$\pm$205&  24.82\\
GN-1089568\tablenotemark{b} & 189.151821 & 62.272229 & 4.05 & 23.2$\pm$1.2&1462$\pm$143 & 24.90\\
GN-1029154\tablenotemark{b} & 189.159025 & 62.260221 & 4.17 & 20.8$\pm$1.5& 2003$\pm$225 & 24.96\\
GN-1086784 & 189.305706 & 62.236946 & 4.40 & 11.7$\pm$1.2 & 3179$\pm$504 & 25.42\\
GN-1008411\tablenotemark{b} & 189.211089 & 62.250271 & 4.41 & \phn 9.4$\pm$1.7 &  3281$\pm$757 &25.44\\
GN-1086855 & 189.286512 & 62.238138 & 4.41 & 16.9$\pm$2.3 & 1724$\pm$271 & 24.61\\
GN-1008671 & 189.161845 & 62.251054 & 4.41 & 22.1$\pm$0.8 & 2273$\pm$95\phn &  24.80\\
GN-1033320\tablenotemark{b} & 189.125779 & 62.287404 & 4.48 & \phn8.2$\pm$1.1 & 1951$\pm$398 &  25.54\\
GN-1085355\tablenotemark{b} & 189.094365 & 62.198974 & 4.88 & 10.7$\pm$0.8 & 1801$\pm$158 & 25.42
\enddata
\hspace{5cm}
\tablenotetext{a}{IDs are from JADES DR2 \citep{Eugenio_2025}.}
\tablenotetext{b}{New broad H$\alpha$ emitters identified in this work. The other emitters are already reported by \citet{Covelo_2024}.}

\end{deluxetable*}

%% file: Tables/Tab2_Fresco.tex
\begin{deluxetable*}{lcccccccc}
\tablewidth{0pt}
\caption{Line-fitting results for the 9 broad H$\alpha$ emitters identified from FRESCO.}
\label{table:Fresco}
\tablehead{
\colhead{JADES ID\tablenotemark{a}} & \colhead{RA} & \colhead{DEC} & \colhead{$z_{\rm spec}$} & \colhead{$f_{\rm H\alpha}$} & \colhead{FWHM$_{\rm H\alpha,~Broad}$} & \colhead{$f_{\text{[O\,{\sc iii}]}\lambda5007}$} &  \colhead{FWHM$_{\text{[O\,{\sc iii}]}}$}
 & \colhead{F444W} \\
\colhead{} & \colhead{(deg)} & \colhead{(deg)} & \colhead{} & \colhead{($10^{-21}$ W/m$^2$)} & \colhead{(km/s)} & \colhead{($10^{-21}$ W/m$^2$)} & \colhead{(km/s)} & \colhead{(mag)} 
}
\startdata
GN-1090253 & 189.285544 & 62.280781 & 5.09 & \phn10.0$\pm$0.7 & 1455$\pm$105 & \nodata & \nodata & 24.65\\
GN-1014406 & 189.072090 & 62.273431 & 5.15 & \phn 13.2$\pm$1.5 & 3212$\pm$639 & \nodata & \nodata& 25.44\\
GN-1034620\tablenotemark{b} & 189.159764 & 62.295924 & 5.19 & \phn21.1$\pm$1.1 & 1077$\pm$71\phn & \nodata&\nodata& 24.31\\
GN-1090549\tablenotemark{b} & 189.235941 & 62.285544 & 5.20 & \phn\phn7.5$\pm$1.0 & 1721$\pm$307&\nodata&\nodata& 25.92\\
GN-9994014 & 189.300125 & 62.212044 & 5.23 & \phn19.5$\pm$1.1& 2084$\pm$156&\nodata&\nodata&24.92\\
GN-1013188 & 189.057100 & 62.268940 & 5.25 & \phn\phn 8.0$\pm$0.6 & 1957$\pm$147 & \nodata&\nodata&25.38\\
GN-1088832 & 189.344282 & 62.263368 & 5.25 & \phn49.7$\pm$1.8 & 2256$\pm$91\phn & \phn8.8$\pm$0.6 &243$\pm$24&23.67\\
GN-1020514 & 189.179302 & 62.292533 & 5.36 & \phn15.5$\pm$0.7 & 1612$\pm$140 & 19.5$\pm$0.4&197$\pm$7\phn&24.79\\
GN-1087388 & 189.281014 & 62.247308 & 5.54 & 130.1$\pm$2.7 & 2965$\pm$55\phn & \phn3.5$\pm$0.3& 150$\pm$23&22.93\\
\enddata
\tablenotetext{a}{IDs are from JADES DR2 \citep{Eugenio_2025}.}
\tablenotetext{b}{New broad H$\alpha$ emitters identified in this work. The other emitters are already reported by \citet{Matthee_2024}.}
\end{deluxetable*}

%% file: 5_Discussion.tex
\section{Discussion} \label{sec: Disc}

\subsection{Luminosity function of broad-line H$\alpha$ emitters} \label{subsec: LF}

We compute the broad H$\alpha$ luminosity function (LF) for the 19 broad H$\alpha$ emitters identified in this work. The broad H$\alpha$ LF is computed following the direct $1/V_{max}$ method \citep{Schmidt_1968}:
\begin{equation}
    \Phi(L) = \frac{1}{d\log L}\sum_i \frac{1}{C_i V_{max,i}}
\end{equation}
where $L$ is the broad H$\alpha$ luminosity of each LF bin, $C_i$ and $V_{max, i}$ represent the completeness correction and the maximum survey volume for the i-th broad H$\alpha$ emitter, respectively. The completeness correction and maximum survey volume are measured following the methodology described in \citet{Lin_2024}. The uncertainties in the LF is estimated via Monte Carlo experiments. Table \ref{table: LF} summarizes the broad H$\alpha$ LF measured in this work.

\begin{figure}[!htp]
    \centering
    \includegraphics[width=0.48\textwidth]{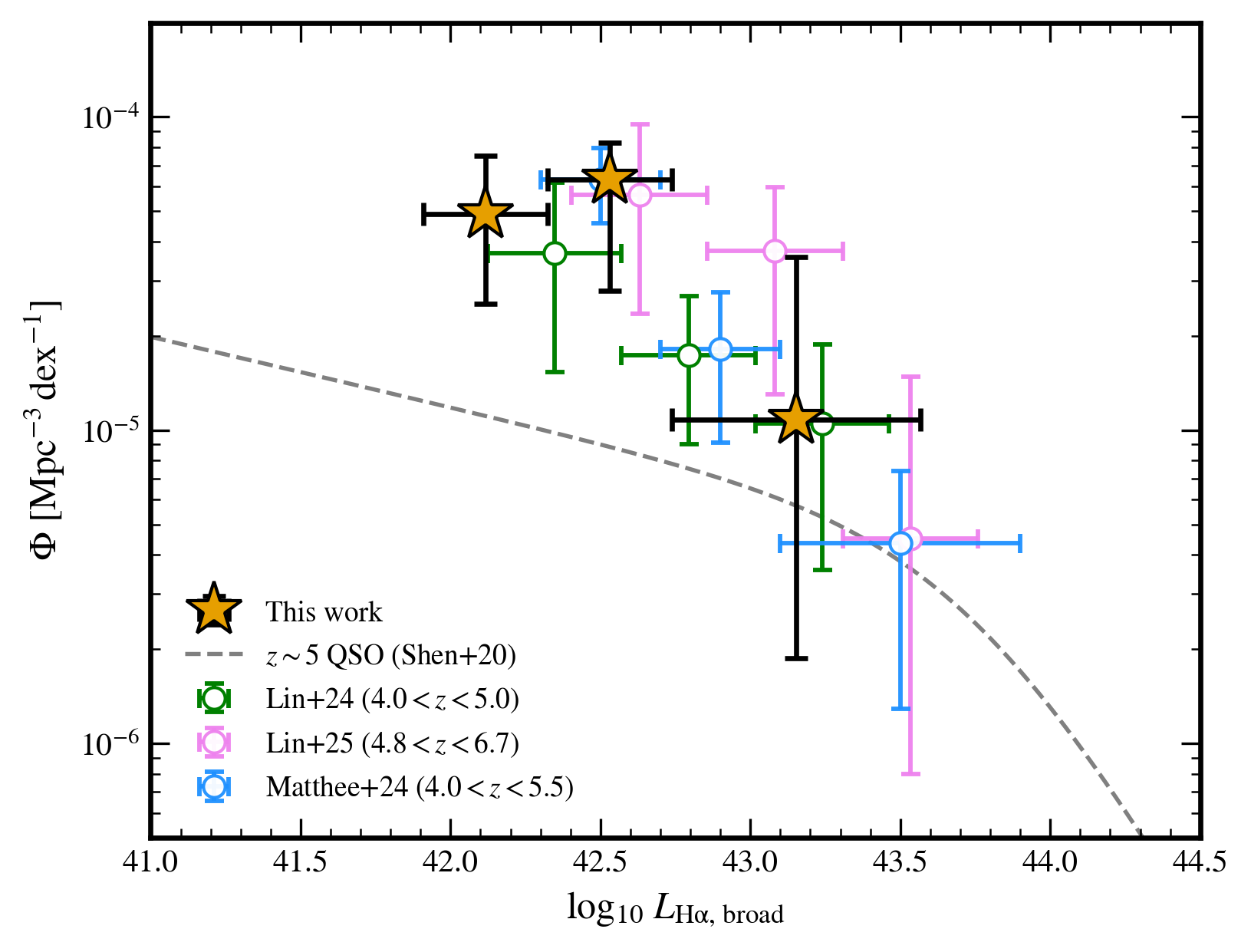}
    \caption{The broad H$\alpha$ LF derived from the broad H$\alpha$ emitter sample identified in this work (yellow stars). For comparison, the broad H$\alpha$ LFs measured in \citet{Matthee_2024}, \citet{Lin_2024}, and \citet{Lin_LRD} are also shown. The gray dot–dashed curve denotes the extrapolation of the quasar bolometric LF at $z \sim 5$ from \citet{Shen_2020}, converted to broad H$\alpha$ luminosity using the local relation between quasar bolometric and broad-line H$\alpha$ luminosities.}
    \label{figure: LF}
\end{figure}

We present the derived broad H$\alpha$ LF of our sample in Figure~\ref{figure: LF}. For comparison, we also include the broad H$\alpha$ LFs derived from JWST-selected broad-line AGN in other studies \citep{Matthee_2024, Lin_2024, Lin_LRD}, as well as the quasar broad H$\alpha$ LF at $z \sim 5$, obtained by converting the bolometric LFs from \citet{Shen_2020}. The conversion assumes the empirical relations between the bolometric luminosity and the AGN-induced H$\alpha$ luminosity from \citet{Greene_2005}, with the bolometric correction from \citet{Richards_2006}. We find that at the faint end (i.e., $L_{\mathrm{H\alpha,broad}} < 10^{43}~\mathrm{erg~s^{-1}}$), the number density of broad H$\alpha$ emitters is significantly higher than the extrapolation of the quasar LF. At the bright end, however, the LF begins to converge toward that of quasars. This trend is consistent with the LFs derived from other JWST-selected AGN in the literature, suggesting a prevalence of broad-line AGN in the early universe. 

\input{Tables/Tab4_LF}

\subsection{The \texorpdfstring{$L_{\rm bol}$--$M_{\rm BH}$}{Lg} Relation}

To gain insights into the accretion efficiency and growth rates of our broad H$\alpha$ emitters,
we estimated the black hole mass ($M_{\rm BH}$) and AGN bolometric luminosity ($L_{\rm bol}$) of our broad H$\alpha$ emitters following the methodology widely adopted in recent studies (e.g., \citealt{ubler_2023,Kocevski_2023,Harikane_2023,Matthee_2024}). More specifically, $M_{\rm BH}$ are estimated using the virial relation proposed by \citet{Reines_2013},
\begin{equation}
\begin{split}
    {\rm log_{10}(M_{\rm BH}/M_{\odot})} = 6.57 + {\rm log}_{10}(\epsilon) + \\ 0.47 \, {\rm log}_{10}(L_{\rm H\alpha, broad}/10^{42}\, {\rm erg \,s}^{-1}) + \\ 2.06 \,  {\rm log}_{10}({\rm FWHM_{broad}}/10^3\, {\rm km \, s^{-1}}),
\end{split}
\end{equation}
where $\epsilon$ represents the scale factor, which we assume to be 1.075, consistent with the value adopted in \citet{Reines_2015}. 

Regarding AGN bolometric luminosity, we follow the approach described in \citet{Harikane_2023}, estimating it from the luminosity of the broad H$\alpha$ component \citep{Greene_2005} and applying the corresponding bolometric correction from \citet{Richards_2006}:
{\small
\begin{equation}
  L_{\rm bol}/10^{44}\, {\rm erg\,s^{-1}} =  10.33(L_{\rm H\alpha, broad}/5.25\times10^{42}{\rm erg\,s^{-1}})^{0.864}
  \label{eqn:lbol}
\end{equation}
}
Table \ref{table:property} presents the estimated values of $M_{\rm BH}$ and $L_{\rm bol}$ for all the broad H$\alpha$ emitters identified in this work. The associated uncertainties are calculated using error propagation. 

\begin{figure*}[!htp]
    \centering
    \includegraphics[width=0.95\textwidth]{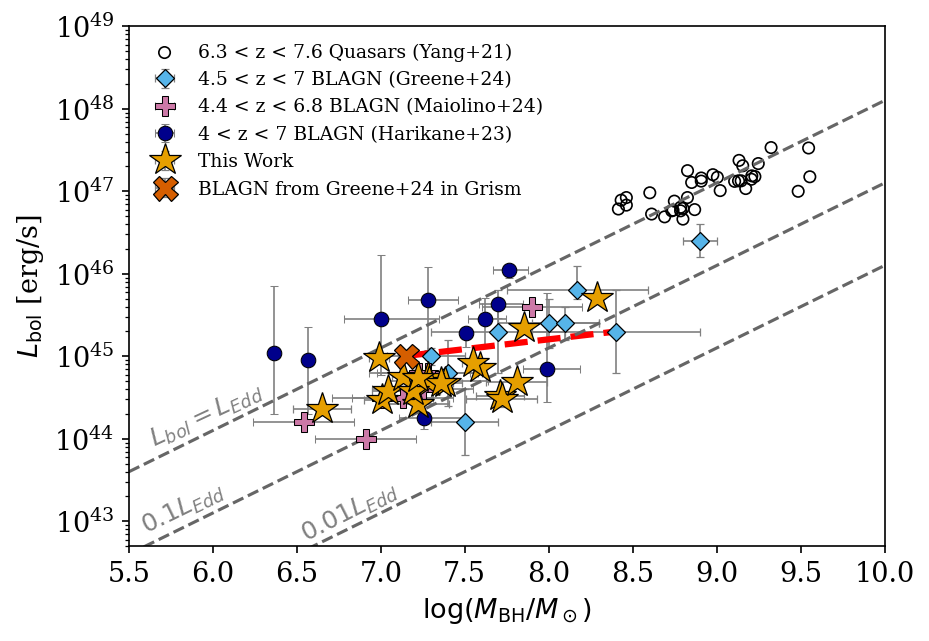}
    \caption{Distribution of 19 broad H$\alpha$ line emitters (yellow stars) on the AGN $L_{\rm bol}$–$M_{\rm BH}$ plane. For comparison, we include BLAGN identified in other studies: \citet{Maiolino_2024a} (violet pluses): \citet{Harikane_2023} (dark blue dots): \citet{Greene_2024} (blue diamonds). We also include UV-selected bright quasars from \citet{Yang_2021} (empty black circles). One of the NIRSpec/prism-detected BLAGN in \citet{Greene_2024} (FWHM\,$\sim$\,4000 km/s) was also observed with NIRCam/grism (SAPPHIRES), but the latter shows a much narrower line width ($\sim$\,1300 km/s; see Appendix \ref{appendix:prism}); its grism-based AGN $L_{\rm bol}$ and $M_{\rm BH}$ are represented by the orange ``X" marker, with the dashed red line indicating how the position shifts between the grism- and the prism-based values. The dashed grey lines represent $L_{\rm bol} = L_{Edd}$, $L_{\rm bol} = 0.1L_{Edd}$, and $L_{\rm bol} = 0.01L_{Edd}$, respectively.}
    \label{figure: LM}
\end{figure*}

We plot the $M_{\rm BH}$ versus H$\alpha$-inferred AGN bolometric luminosities in Figure \ref{figure: LM}. For comparison, we include UV-selected luminous quasars at $z > 6.3$ from \citet{Yang_2021}, broad-line AGN (BLAGN) identified with NIRSpec grating spectroscopy at $z\approx 4.4–6.8$ from \citet{Maiolino_2024a}\footnote{We exclude three sources from \citet{Maiolino_2024a} that exhibit two broad components.}, BLAGN at $z\approx4–7$ from \citet{Harikane_2023}, and BLAGN detected with NIRSpec prism spectroscopy at $z\approx4.5–7$ from \citet{Greene_2024}.  Notably, the $M_{\rm BH}$ for those BLAGN were derived using the same methodology employed in this study, whereas \citet{Yang_2021} estimated $M_{\rm BH}$ based on the continuum luminosity at 3000 \AA~and the FWHM of the \ion{Mg}{2} line.

The majority of our sample is found to be accreting at an Eddington ratio of the order of 0.1, lower than that of UV-selected luminous quasars, consistent with the intrinsically faint nature of our broad H$\alpha$ emitters. These moderate accretion rates are in line with those reported for BLAGN in \citet{Maiolino_2024a} and \citet{Greene_2024}. In contrast, the bolometric luminosities of the BLAGN in \citet{Harikane_2023} are derived using a more complex methodology, where $L_{\rm bol}$ as estimated from Equation~\ref{eqn:lbol} is treated as a lower limit. Taking this into account, we find no significant difference in the inferred accretion rates between our sample and that of \citet{Harikane_2023}.

Interestingly, we notice that the BLAGN identified with NIRSpec prism($R\sim100$) spectroscopy in \citet{Greene_2024} tend to occupy the high-mass end of the $M_{\rm BH}$ distribution relative to our sample. The recent JWST Cycle 3 General Observer program Slitless Areal Pure-Parallel High-Redshift Emission Survey (SAPPHIRES-PID: 6434; PI: E. Egami; \citealt{Sun_2025}) observed one BLAGN (FWHM\,$\sim$\,4000\,km/s) from \citet{Greene_2024} with the F444W grism, providing an independent H$\alpha$ line measurement. While the source still exhibits a broad feature, the line width appears significantly narrower (FWHM\,$\sim$1300\,km/s). Given the grism’s higher spectral resolution, we adopt the latter measurement and mark the resulting $L_{\rm bol}$ and $M_{\rm BH}$ with an ``X” in Figure~\ref{figure: LM}, which suggests the possibility that some of the line widths measured by NIRSpec/prism may be overestimated. The details about the analysis of the SAPPHIRES spectrum are presented in Appendix~\ref{appendix:prism}

\subsection{The \texorpdfstring{$M_{\rm BH}$-$M_{*}$}{Lg} Relation} \label{subsec: SED}
In this work, the stellar masses ($M_{*}$) of the AGN hosts are measured by SED-fitting using the \texttt{Prospector} Bayesian inference framework (\citealt{Leja_2017,  2021ApJS..254...22J, ben_johnson_2022_6192136}) based on the Flexible Stellar Population Synthesis (FSPS; \citealt{Conroy_2009, Conroy_2010}) code. To achieve reliable AGN identification, we adopt a modified version of \texttt{Prospector} presented in \citet{Lyu_2024}, which incorporates a semi-empirical model for AGN UV-to-mid-IR continua with nebular emission lines and dust attenuation and replaces the default AGN torus model in \texttt{Prospector}. 

\begin{figure}[!htp]
    \centering
    \includegraphics[width=0.45\textwidth]{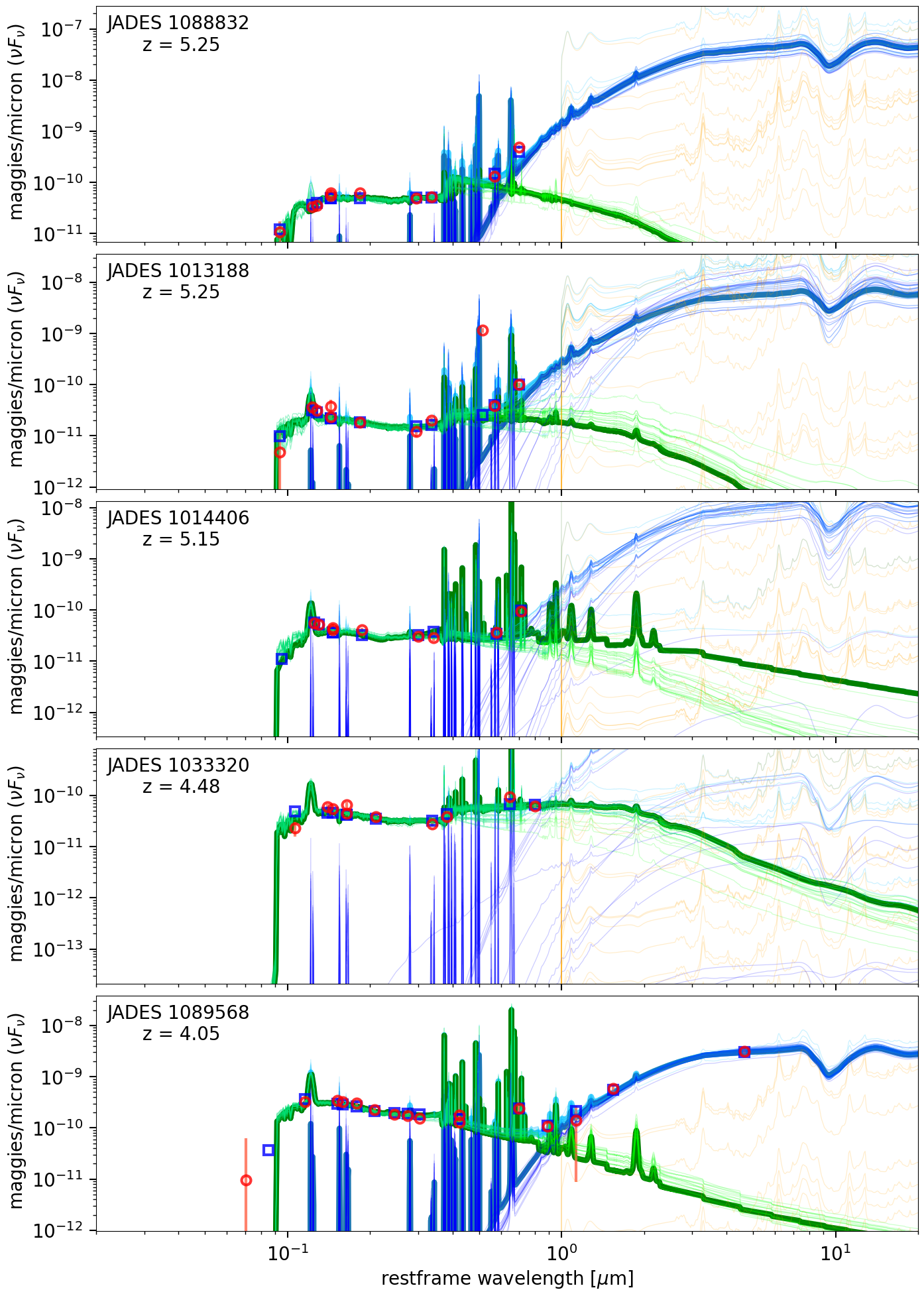}
    \caption{Example SED fittings of our sample: the green line is the stellar component, the blue line is the AGN model and the orange line is the galaxy dust emission. The measurements are shown as the open red circles with error bars, and the blue squares are the result of synthetic photometry on the model. The thin lines are showing 20 randomly selected models from the posterior.}
    \label{figure: SED}
\end{figure}

During the SED-fitting process, we include not only the deep JWST/NIRCam photometry but also multi-band data spanning the optical to near-infrared from HST/ACS and HST/WFC3, as well as longer-wavelength infrared data from Spitzer, if they are available. Examples of galaxy SED-fitting results are shown in Figure~\ref{figure: SED}. The bottom panel presents the only source with a robust detection at longer wavelengths from Spitzer. For the remaining sources, due to the limited wavelength coverage of the NIRCam photometry, clear AGN signatures are identified through SED analysis in only 9 out of 18 galaxies, despite the presence of broad H$\alpha$ emission lines. Therefore, follow-up MIRI observations to obtain mid-infrared data for these broad H$\alpha$ emitters would be valuable for better constraining their physical properties. Table \ref{table:property} lists the best-fit values and corresponding uncertainties of $M_{*}$ for all broad H$\alpha$ emitters. Table \ref{table:property} also summarizes the key physical properties of the 19 broad H$\alpha$ emitters. 

\begin{figure*}[!ht]
    \centering
    \includegraphics[width=0.95\textwidth]{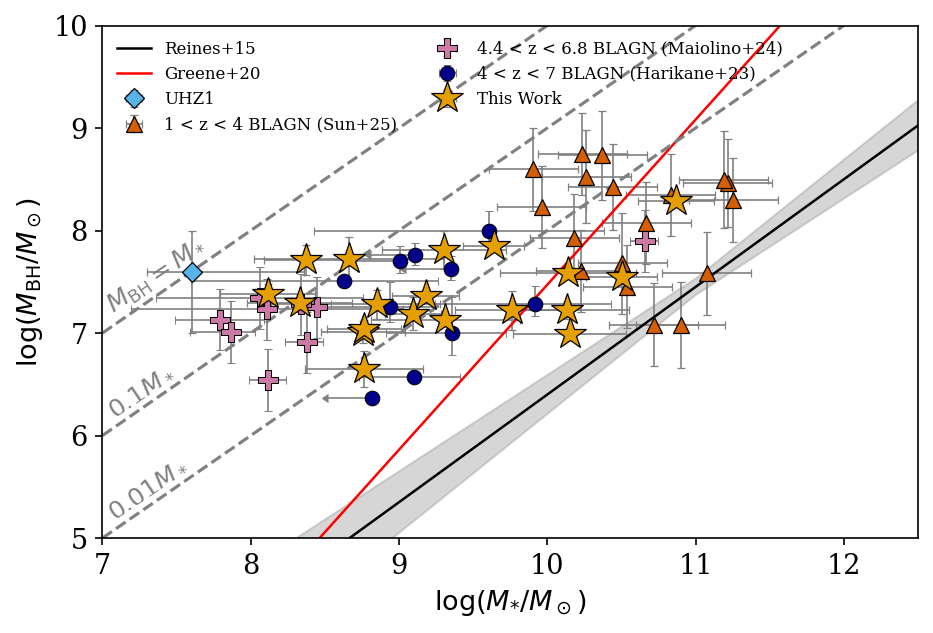}
    \caption{$M_{\rm BH}$–$M_{*}$ relation for the broad H$\alpha$ emitter sample (yellow stars). For comparison, we include BLAGN samples spanning a wide redshift range from the literature: \citet{Maiolino_2024a} (violet pluses), \citet{Harikane_2023} (dark blue dots), and \citet{Suny_2025b} (red triangles). One potential $z>10$ AGN, UHZ1 (blue diamond), is also shown. The solid black line and shaded gray region represent the local $M_{\rm BH}$–$M_{*}$ relation derived by \citet{Reines_2015}, while the solid red line shows the local relation (all, limits case) from \citet{Greene_2020}.}
    \label{figure: MM}
\end{figure*}

Figure \ref{figure: MM} illustrates the distribution of broad H$\alpha$ emitters identified in this study on the $M_{\rm BH}-M_{*}$ plane. To further explore the evolution of the $M_{\rm BH}-M_{*}$ relation over cosmic time, we incorporate AGN recently discovered with JWST across a wide redshift range. These include one potential AGN at $z > 10$, UHZ1 \citep{Bogdan_2024}, as well as BLAGN in the GOODS-N field at $4.4 < z < 6.8$ \citep{Maiolino_2024a} and $1 < z < 4$ \citep{Suny_2025b}, and BLAGN at $4 < z < 7$ identified by \citet{Harikane_2023}. The local $M_{\rm BH}-M_{*}$ relations for AGN hosts derived in \citet{Reines_2015} (hereafter RV15) and \citet{Greene_2020} (all, limits case, hereafter GSH20) are also included for comparison purposes. 

We find that nearly all AGN recently revealed by JWST observations at $z > 4$ lie above RV15, with the majority located significantly above the relation. When compared to the GSH20, which is considered a more robust local relation that shows no bias with galaxy type and provides a better match to broad-line selected AGN at $1 < z < 4$ \citep{Suny_2025}, the offset is somewhat reduced. Nevertheless, many sources still exhibit higher $M_{\rm BH}$/$M_{*}$ ratios than those predicted by GSH20. The over-massive $M_{\rm BH}$ relative to the $M_{*}$ of host galaxy, when compared with the local scaling relations, is consistent with findings from numerous previous studies of AGN at similar redshifts (e.g., \citealt{Kocevski_2023, ubler_2023}). These results may suggest a different evolution path of the $M_{\rm BH}$–$M_{*}$ relation in the early Universe compared to that observed locally. 
Additionally, the presence of SMBHs with $M_{\rm BH}$ above the local scaling relations aligns with theoretical predictions from models invoking heavy black hole seeds and super-Eddington accretion, which are proposed to explain the rapid growth of SMBHs at very early time \citep{Trinca_2022, Volonteri_2023, Schneider_2023}. Alternatively, the growth of the host galaxy can also be suppressed by strong feedback processes (see \citealt{Zhu_2025} and references therein). However, it is also important to note that, although our AGN sample is selected based on the presence of broad emission lines in spectroscopy, selection bias still exists to some extent. Specifically, due to the detection limits of the grism observations and the requirement of FWHM $>$ 1000 km/s, we may fail to detect very faint, low-mass AGN (see Section \ref{subsec: Detectability} for more details), which could otherwise help fill the observed gap between our sample and the local $M_{\rm BH}$-$M_{*}$ relation.

\subsection{Are Broad Line Emitters LRDs?} \label{sec: LRD}
In this section, we evaluate the fraction of our spectroscopically confirmed broad H$\alpha$ emitters that can be photometrically classified as LRDs by applying the selection criteria outlined in \citet{Pier_2024}:
\begin{itemize}
    \item Blue slope: F150W $-$ F200W $<$ 0.8 mag
    \item Red slope: F277W $-$ F444W $>$ 0.7 mag
\end{itemize}
Since the compactness criterion was already applied during the selection of broad H$\alpha$ emitters, and the possibility of these sources being brown dwarfs was excluded, we only implemented color cuts for our sample. Of the 19 sources in our sample, only 9 are covered by JADES imaging, and thus have available photometry in the F150W, F200W, and F277W bands. For the remaining 10 sources, we use \texttt{Synphot} to estimate their photometry in these bands based on their best-fit SEDs. 

\begin{figure*}[htp]
    \centering
    \begin{minipage}{0.47\textwidth}
        \centering
        \includegraphics[width=\linewidth]{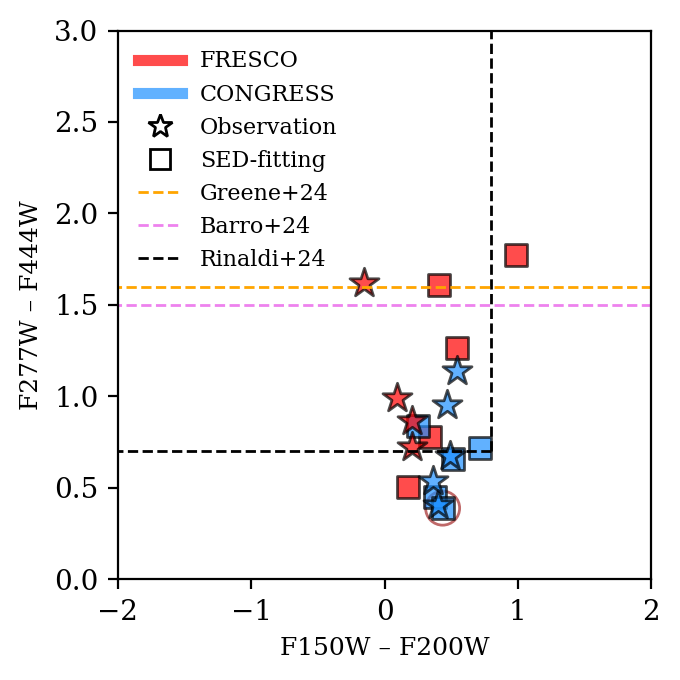}
    \end{minipage}
    \begin{minipage}{0.47\textwidth}
        \centering
        \includegraphics[width=\linewidth]{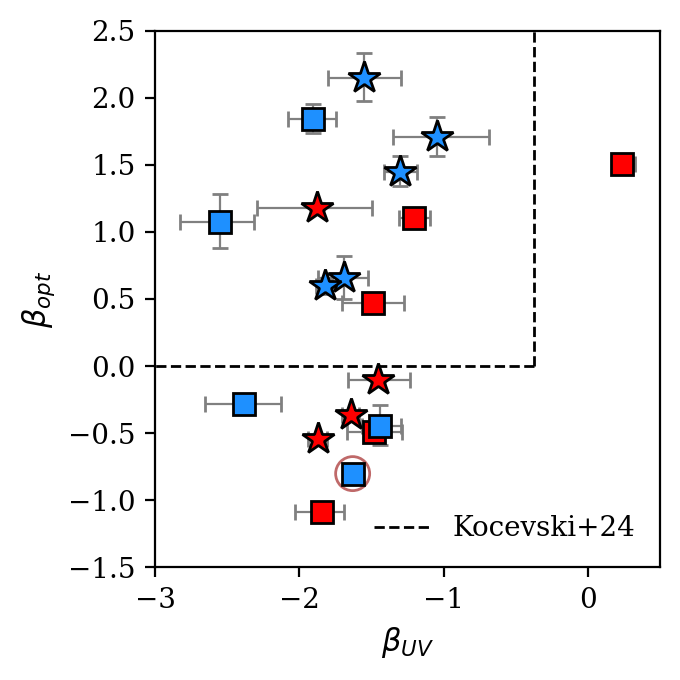}
    \end{minipage}
    \caption{Left panel: Distribution of broad H$\alpha$ emitters on the color–color diagram (Red: FRESCO; Blue: CONGRESS). Stars indicate sources with available photometric measurements in the F150W, F200W, and F277W bands, while squares represent sources for which photometry in these bands was estimated from best-fit SEDs. The dashed black lines indicate the LRD selection criteria defined by \citet{Pier_2024}, while the dashed orange and violet lines represent LRD selection criteria proposed by \citet{Greene_2024} and \citet{Barro_2024}, respectively, which have shown high yield selection of AGN. Right panel: Distribution of our sample in the $\beta_{opt}-\beta_{UV}$ plane. The dashed black lines indicate the LRD selection criteria adopted in \citet{Kocevski_2024}. The brown circle in each panel highlights the broad H$\alpha$ emitter with mid-infrared detections from Spitzer.}
    \label{figure: color}
\end{figure*}

The left panel of Figure \ref{figure: color} shows the positions of the broad H$\alpha$ emitters on the color-color plot. We find that a significant fraction of our sample ($\sim$42\%) cannot be photometrically selected as LRDs, with the majority (7 out of 8) failing to meet the red slope criterion. This result is consistent with \citet{Kevin_2024}, who applied a stricter red slope threshold and found that only 30\% of spectroscopically identified BLAGN with JWST meet the LRD selection criteria. These findings suggest that not all broad-line emitters with a compact, red appearance exhibit the characteristic ``V-shaped" SED of LRDs, and that many lack a steep red slope.

Additionally, since bright H$\alpha$ emission is observed in the F444W for all broad H$\alpha$ emitters identified in FRESCO, applying a relatively loose red slope criterion raises concerns about those sources that satisfy the LRD color selection. It remains unclear whether these sources intrinsically exhibit such steep red slopes, or whether their observed colors are primarily driven by line-boosting effects \citep{Kevin_2024}.

A more conservative approach to selecting LRDs with a high AGN yield through color cuts involves applying a higher cut on the red slope, in order to reduce contamination from line-boosting effects, as proposed by \citet{Barro_2024} and \citet{Greene_2024}. In the left panel of Figure~\ref{figure: color}, we add the selection criteria proposed in both studies:
\begin{itemize} 
\item Red slope: F277W $-$ F444W $>$ 1.5 mag \citep{Barro_2024}
\item Red slope: F277W $-$ F444W $>$ 1.6 mag \citep{Greene_2024}
\end{itemize}
However, even without considering the blue slope requirement, applying these stricter thresholds yields only 3 LRD candidates in both cases, missing 8 LRDs identified by our original selection criteria and 84\% of our broad-line emitters. Therefore, the trade-off between completeness and purity in color-based LRD selection should be taken into consideration when determining the appropriate color cuts to apply.

In addition to color-based selection, \citet{Kocevski_2024} proposed an alternative approach for identifying LRDs by applying cuts on the UV and optical continuum slopes, derived from fitting photometry across multiple bands blueward and redward of the Balmer break at 3645\AA. This method enables the detection of LRDs with reduced contamination from galaxies that exhibit strong Balmer breaks but lack a rising red continuum.  To make optimal use of the available photometry, we derive the UV continuum slope using the F090W, F115W, and F182M bands. For sources at redshift $z<4.75$, we use F210M, F277W, and F356W to measure the optical continuum slope, while for higher-redshift sources (i.e., $z\geq 4.75$), we instead use F277W, F356W, and F444W. For sources lacking photometry in F277W, we estimate their fluxes in F277W from their best-fit SEDs. The uncertainties on the measured slopes are estimated using a Monte Carlo approach following the methodology described in \citet{Kocevski_2024}.

We adopt the same cuts used in \citet{Kocevski_2024}:
\begin{itemize}
    \item Blue slope: $\beta_{\rm UV} <$ -0.37 
    \item Red slope: ~$\beta_{\rm opt} >$ 0 
\end{itemize}
The right panel of Figure~\ref{figure: color} shows the distribution of our sample in the $\beta_{\rm opt}-\beta_{\rm UV}$ plane. Based on this selection approach, 47\% of our sample do not satisfy the selection criteria. Similarly, most of the outliers fail to satisfy the red-slope criterion. All three LRD selection techniques miss an important fraction of our broad-line emitters (i.e., 42\%, 84\%, and 47\%) suggesting that current LRD selection methods may overlook a substantial fraction of the broad-line emitter population. 

However, most of the broad H$\alpha$ emitters that failed the LRD color selection do not satisfy the red slope criterion, with the majority originating from CONGRESS. In these cases, the [\ion{O}{3}] doublet falls within the F277W filter, making the F277W $-$ F444W color less red, and causing these sources to fall outside the selection boundary. An example of an object that clearly exhibits LRD-like behavior in its SED but is not color-selected as an LRD due to strong [\ion{O}{3}] emission is presented in \citet{Pier_2025}. This further highlights the challenges and limitations of color-based LRD selection at these redshifts.

While several alternative color selection criteria of LRDs have been proposed in the literature (e.g., \citealt{Greene_2024, Pablo_2024}), we adopt the criteria used by \citet{Pier_2024} in this work to maintain consistency, as their LRD sample is also constructed in the GOODS-N field.

\begin{figure}[htbp!]
    \centering
    \includegraphics[width=0.47\textwidth]{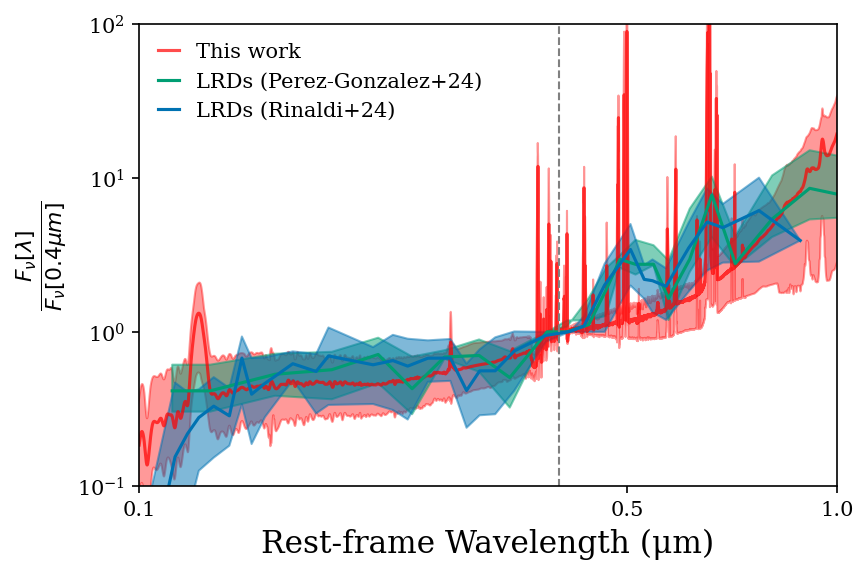}
    \caption{Stacked SEDs for the broad-line emitter sample identified in this work (red) and for the LRD samples selected in \citet{Pier_2024} (dark blue) and \citet{Pablo_2024} (teal green), normalized at a rest-frame wavelength of 0.4\,$\mu$m. Shaded regions denote the 1-$\sigma$ uncertainties on the average SEDs.}
    \label{figure: Average SED}
\end{figure}

To further explore the relationship between broad-line emitters and photometrically selected LRDs, we perform a comparison between the average SEDs of our broad H$\alpha$ emitters and the LRD sample selected in \citet{Pier_2024}.  We follow the methodology outlined in \citet{Pablo_2024} to derive the average SED of the LRD sample by stacking the SEDs of LRDs with available spectroscopic redshifts (including their emission lines) and computing the mean flux for every ten consecutive data points in the wavelength-ordered list. In contrast, the average SED of our broad H$\alpha$ emitter sample is constructed using the best-fit SEDs obtained in Section \ref{subsec: SED}, as about half of the sources in our sample have photometry available in only 6 filters. Figure \ref{figure: Average SED} displays the average rest-frame SEDs for both broad H$\alpha$ emitters and LRDs, with flux normalized at 0.4\,$\mu$m. The average SED of LRDs selected in the GOODS-S field presented in \citet{Pablo_2024} is also included for comparison.

Overall, we do not observe significant differences between the average SEDs of our broad H$\alpha$ emitters and LRDs within the rest-frame 0.1–1\,$\mu$m range. The SEDs closely align up to rest-frame 0.4\,$\mu$m, suggesting that LRDs may share similar UV emission properties with broad-line emitters. However, beyond 0.5\,$\mu$m, the average SEDs of LRDs exhibit a noticeably steeper slope toward longer wavelengths compared to those of broad-line emitters, with particularly significant deviations at 0.5 and 0.65\,$\mu$m. Given that the observed continuum in the average LRD SEDs between these wavelengths is comparable to that of broad-line emitters, we suspect that the steep red slope is not an intrinsic feature of at least some LRDs, but rather a result of line-boosting effects from strong emission lines, such as [\ion{O}{3}] and H$\alpha$, falling within individual wide filters \citep{Kevin_2024}. Figure \ref{figure: Average SED} further supports the concern that adopting a low red slope cut for LRD selection may lead to contamination by sources that exhibit ``fake" steep red slopes induced by such line-boosting effects, particularly when there are strong emission lines falling within the filters used for color selection.

\subsection{Properties of Non-LRD AGN Galaxies}

It should be noted that BLAGN galaxies that do not satisfy the NIRCam-based red-slope criterion (and therefore drop out the LRD selection) could still exhibit an SED upturn (i.e., a V-shaped SED) at $\lambda_{\rm obs}$\,$>$\,5 \micron\ (i.e., beyond the NIRCam wavelength coverage).  A good example is JADES 1089568 at $z$\,$=$\,4.05 shown in the bottom panel of Figure~\ref{figure: SED}.  The figure shows that this object has a very blue color over the NIRCam passbands (e.g., due to its star-forming host), and in fact it is one of the bluest objects in Figure~\ref{figure: color} in terms of the F277W-F444W color and $\beta_{\rm opt}$ slope.  And yet it clearly exhibits a red SED rising beyond $\lambda_{\rm obs}$\,$\sim$\,5\,\micron\ ($\lambda_{\rm rest}$\,$\gtrsim$\,1\,\micron), which can be fit well with our AGN SED model.  Such an SED is actually typical of AGN galaxies and widely seen across the redshift \citep[e.g.,][]{Lyu_2024} 
with hot dust emission from the AGN torus starting to dominate only in the rest-frame near-/mid-infrared.
This illustrates why the LRD selection (and more generally NIRCam-only color selections) is highly incomplete when selecting AGN galaxies (see also \cite{Kevin_2024}).


\subsection{Detectability of Broad H\texorpdfstring{$\alpha$}{Lg} Line in Grism Spectra} \label{subsec: Detectability}

We identify 19 photometrically selected LRDs from \citet{Pier_2024} that lie within the FRESCO and CONGRESS fields and have H$\alpha$ emission detected in the grism spectra. However, 68\% of these sources (13 out of 19) do not show broad H$\alpha$ line profiles (i.e.,FWHM $<$ 1000 km/s) in the grism spectroscopy, with the exception of the 6 broad H$\alpha$ emitters identified in this work. Such a low fraction of LRDs displaying broad-line features is consistent with the findings reported by \citet{Pablo_2024}.

Although no broad-line feature is observed in these LRDs, we cannot fully rule out the possibility that they are broad-line emitters, as the broad component could be hidden due to the insufficient sensitivity of the grism spectra. Notably, two sources that show no broad line feature in the grism spectra have been reported to exhibit broad line features in NIRSpec medium-resolution grating data in previous work \citep{Maiolino_2024a}. To illustrate how S/N may affect the detection of broad components, we present both the NIRSpec/grating and NIRCam/grism spectra for these two sources in Appendix \ref{appendix:grating}.

Figure \ref{figure: Mag} presents the distribution of our broad-line emitters and the photometrically selected LRDs of \citet{Pier_2024}, which lack broad-line features in the grism spectra, on the redshift-magnitude plane. Overall, our broad-line emitters are significantly brighter than the LRDs, with a division at $\sim$26 mag that effectively separates the two populations. Notably, the two sources that exhibit broad-line features in the grating spectra but not in the grism are both fainter than 26 mag, suggesting that some faint LRDs may indeed be broad-line emitters, but their broad components are too faint to be detected in grism observations. 
In other words, the broad-line search with the CONGRESS/FRESCO data is effective only down to $\sim$\,26\,mag.

\begin{figure}[!htp]
    \centering
    \includegraphics[width=0.48\textwidth]{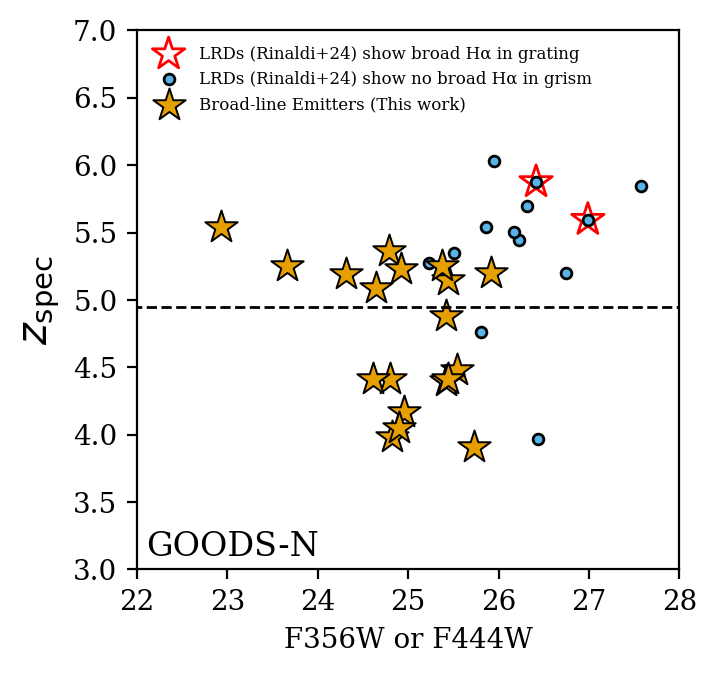}
    \caption{Distribution of broad H$\alpha$ emitters (yellow stars) and photometrically selected LRDs (blue dots) from \citet{Pier_2024} that show no broad H$\alpha$ feature in grism on the redshift–magnitude plane. The two blue dots enclosed by red stars represent LRDs exhibiting broad H$\alpha$ emission in grating spectra but lacking such features in grism spectra. Sources below the dashed line have their H$\alpha$ emission falling into the F356W band, so their F356W magnitudes are shown. Similarly, sources above the dashed line have their H$\alpha$ emission in the F444W band, so their F444W magnitudes are shown.}
    \label{figure: Mag}
\end{figure}

\input{Tables/Tab3_property}

%% file: Tables/Tab4_LF.tex
\begin{deluxetable}{cccc}
\tabletypesize{\footnotesize}
\tablewidth{0pt}
\caption{Broad H$\alpha$ Luminosity Function of Our Broad H$\alpha$ Emitter Sample.}
\label{table: LF}
\tablehead{
\colhead{$\log L_{\rm H\alpha,broad}$} & \colhead{$\Delta\log L_{\rm H\alpha,broad}$} & \colhead{N} & \colhead{$\Phi (\rm 10^{-5}Mpc^{-3}dex^{-1})$}}
\startdata
42.117 & 0.414 & 7 & $4.891^{+2.634}_{-2.362}$ \\
42.531& 0.414 & 10 & $6.312^{+1.963}_{-3.523}$ \\
43.153 & 0.819 & 2 & $1.080^{+2.483}_{-0.893}$ \\
\enddata
\end{deluxetable}

%% file: Tables/Tab3_property.tex
\begin{deluxetable*}{lccccc}
\tablewidth{\textwidth}
\label{table:property}
\caption{Physical properties of the broad H$\alpha$ emitters}
\tablehead{
\colhead{JADES ID} & \colhead{$L_{\text{H}_{\alpha},~ \text{broad}}$} & \colhead{$\log(M_{\rm BH}/M_{\odot})$} & \colhead{$L_{\text{bol}}$} & \colhead{$L_{\rm H\alpha,~broad}/L_{\rm H\alpha,~tot}$} & \colhead{$\log{(M_{*}/M_{\odot})}$} \\
\colhead{} & \colhead{($10^{42}$ erg/s)} & \colhead{} & \colhead{($10^{44}$ erg/s)} & \colhead{} & \colhead{}}
\startdata
GN-1014406 &  \phn2.21 $\pm$ 0.38 &  7.81 $\pm$ 0.18 & \phn4.88 $\pm$ 0.73 &  0.60 $\pm$ 0.05 &  \phn 9.30 $\pm$ 0.42 \\
GN-1085355  &                                \phn2.22 $\pm$                        0.17 &         7.29 $\pm$        0.08 &                         \phn4.91 $\pm$                 0.33 &   0.84 $\pm$         0.04 & \phn8.33 $\pm$ 0.35\\
GN-1087388 &                                  32.90 $\pm$                        0.76 &         8.29 $\pm$        0.02 &                        50.50 $\pm$                 1.00 &   0.76 $\pm$         0.01 & 10.87 $\pm$ 0.26\\
GN-1088832 &                                    12.70 $\pm$                       0.47 &         7.85 $\pm$        0.04 &                        22.20 $\pm$                 0.71 &   0.88 $\pm$         0.01 & \phn 9.64 $\pm$ 0.21\\
GN-1008671 &                                    \phn3.49 $\pm$                        0.13 &         7.59 $\pm$        0.04 &                         \phn7.26 $\pm$                 0.24 &   0.81 $\pm$         0.01 & 10.14 $\pm$ 0.46\\
GN-1013188 &  \phn 2.14 $\pm$ 0.16 & 7.36 $\pm$ 0.07 & \phn4.75 $\pm$ 0.31 & 0.92 $\pm$ 0.02 & \phn 9.18 $\pm$ 0.23\\
GN-1020514 &                                  \phn2.54 $\pm$                        0.20 &         7.22 $\pm$        0.08 &                         \phn5.52 $\pm$                 0.38 &   0.53 $\pm$         0.02 & 10.14 $\pm$ 0.42\\
GN-1029154 & \phn2.16 $\pm$ 0.21& 7.38 $\pm$ 0.10 & \phn4.79 $\pm$ 0.39 & 0.61 $\pm$ 0.04 & \phn 8.12 $\pm$ 0.32\\
GN-1033320 &                \phn1.09 $\pm$                        0.19 &         7.22 $\pm$        0.19 &                         \phn2.65 $\pm$                 0.41 &   0.66 $\pm$         0.06 & \phn 9.76 $\pm$ 0.38\\
GN-1034620 &             \phn4.83 $\pm$                        0.26 &         6.99 $\pm$        0.06 &                         \phn9.62 $\pm$                 0.44 &   0.81 $\pm$         0.02 & 10.15 $\pm$ 0.38\\
GN-1082263 &                \phn0.91 $\pm$ 0.16 &         6.65 $\pm$ 0.17&                         \phn2.28 $\pm$ 0.33&   0.63 $\pm$ 0.07 & \phn8.77 $\pm$ 0.40 \\
GN-1086784 & \phn1.41 $\pm$ 0.21& 7.71 $\pm$ 0.15 & \phn3.31 $\pm$ 0.43 &   0.62 $\pm$ 0.04 & \phn8.38 $\pm$ 0.29\\
GN-1086855 & \phn2.52 $\pm$ 0.36 & 7.28 $\pm$ 0.14& \phn5.48 $\pm$ 0.69 & 0.77 $\pm$ 0.06 & \phn8.85 $\pm$ 0.43\\
GN-1087315 &                 \phn1.22 $\pm$                        0.13 &         7.01 $\pm$        0.11 &                         \phn2.92 $\pm$                 0.26 &   0.67 $\pm$         0.04 & \phn8.76 $\pm$ 0.28\\
GN-1089568 &          \phn1.66 $\pm$                        0.16 &         7.04 $\pm$        0.09 &                         \phn3.83 $\pm$                 0.32 &   0.45 $\pm$         0.03 & \phn 8.76 $\pm$ 0.25\\
GN-1090253 &               \phn2.60 $\pm$                        0.18 &         7.13 $\pm$        0.07 &                         \phn5.63 $\pm$                 0.34 &   0.96 $\pm$         0.02 & \phn 9.31 $\pm$ 0.50\\
GN-1090549 &             \phn1.62 $\pm$                        0.25 &         7.19 $\pm$        0.16 &                         \phn3.75 $\pm$                 0.49 &   0.76 $\pm$         0.06 & \phn 9.09 $\pm$ 0.19\\
GN-1008411 & \phn1.28 $\pm$ 0.32 &7.72 $\pm$ 0.21 & \phn3.06 $\pm$ 0.66 & 0.70 $\pm$ 0.06 & \phn 8.66 $\pm$ 0.64\\
GN-9994014 &             \phn4.08 $\pm$                        0.30 &         7.55 $\pm$   0.07 &                        \phn8.31 $\pm$                 0.54 &   0.72 $\pm$         0.02 & 10.50 $\pm$ 0.24\\
\enddata
\end{deluxetable*}

%% file: 6_Conclusion.tex
\section{Conclusion} \label{sec:highlight}
In this study, we conduct a comprehensive spectroscopic search for broad H$\alpha$ emitters at $z\approx3.7 - 6.5$ in the GOODS-N field with JWST/NIRCam WFSS and imaging data. By combining photometric data from JADES with slitless spectroscopic observations from CONGRESS and FRESCO, we identify 19 broad H$\alpha$ emitters at $z\approx 3.9-5.5$ with FWHM $>$ 1000 km/s, including 9 newly identified sources. 

\begin{itemize}
    \item Among the 19 broad H$\alpha$ emitters, 18 of them have a broad component that contributes more than 50\% of the total H$\alpha$ flux. Additionally, [\ion{O}{3}]~$\lambda5007$ emission line is detected in three sources, with line widths significantly narrower than those of the broad H$\alpha$ components, suggesting that the observed broad features of these sources are mainly driven by the AGN BLR. By computing the broad H$\alpha$ LF of our sample, we find that it is consistent with those of other JWST-selected BLAGN in the literature.
    
    \item We derived black hole masses and AGN bolometric luminosities for our sample based on the measured FWHM and line fluxes of the broad H$\alpha$ components. Most sources are found to be accreting at $\sim$10\% of the Eddington limit, consistent with the accretion rates of BLAGN selected with NIRSpec in other studies.
    
    \item The stellar masses of the host galaxies in our sample were derived through SED-fitting using \texttt{Prospector}. Consistent with previous studies, our sample exhibits higher $M_{\rm BH}/M_*$ ratios relative to the local $M_{\rm BH}-M_*$ relations reported by \citet{Reines_2015} and \citet{Greene_2020}. However, given that our sample is selected based on the detection of broad H$\alpha$ emission, some degree of selection bias may be present. In particular, low $M_{\rm BH}$ sources may be missed due to the limited sensitivity of the grism, which could prevent the detection of faint broad-line components.
    
    \item By testing various LRD selection techniques proposed in the literature, we find that all of these methods will miss a significant fraction of our spectroscopically confirmed broad H$\alpha$ emitters, while the majority of the broad H$\alpha$ emitters that failed the LRD color selection lack the steep red slopes. This suggests that not all broad-line emitters with compact and red appearance exhibit the characteristic ``V-shaped" SEDs of LRDs.
    
    \item To further investigate the nature of LRDs, we compared the average SED of our broad-line emitter sample with that of the photometrically selected LRDs from \citet{Pier_2024} in the same field. The average SEDs are largely consistent with each other, except for a steeper red slope in the LRD sample at rest-frame wavelength $>$ 0.5\,$\mu$m, likely due to the contribution of [\ion{O}{3}] and H$\alpha$ emission. Given the much lower continuum levels observed between these two lines, we conclude that the red slope, at least in some LRDs, would be naturally less steep if emission lines are removed, as discussed in \citet{Kevin_2024}.
    
    \item We further examined the LRDs in the GOODS-N field with H$\alpha$ detections in the grism spectra and find that $\sim$68\% do not exhibit broad H$\alpha$ features. This suggests that the LRD population may not only consist of broad-line emitters. Additionally, we identified two LRDs that lack broad H$\alpha$ features in the NIRCam grism spectra but do show broad components in the NIRSpec grating spectra. Both of these sources are fainter than 26 mag in the filter where H$\alpha$ is observed, whereas the majority of our broad H$\alpha$ emitters are brighter than this threshold. This implies that the broad-line search with the CONGRESS/FRESCO data is effective only down to $\sim$\,26\,mag, and therefore that some LRDs may indeed host broad-line components that are undetected in the grism data due to its limited sensitivity.   
\end{itemize}

This work identifies an abundant population of broad H$\alpha$ emitters in the GOODS-N field at $3.9 < z < 5.5$, providing a valuable sample for studying the abundance and properties of broad H$\alpha$ emitters in the early Universe. Our results also highlight the strong selection capabilities of grism spectroscopy and offer insights into the brightness threshold required for detecting broad H$\alpha$ emitters with grism. Future follow-up observations with NIRSpec will be crucial for confirming the nature of these sources and further characterizing this population.